\documentclass[notitlepage,nofootinbib,showpacs,unsortedaddress,aps,prd,twocolumn]{revtex4-1}
\usepackage[utf8x]{inputenc}
\usepackage{amsmath, graphicx, amsfonts, slashed}
\usepackage{hyperref}

\newcommand{\com}[2]{\ensuremath{\left[ #1 , #2 \right] }}
\newcommand{\acom}[2]{\ensuremath{\left\{ #1 , #2 \right\}}}
\newcommand{\cp}[3]{\ensuremath{\left( #1\! \times\! #2 \right)^{#3} }}
\newcommand{\cpt}[3]{\ensuremath{\left( #1 \,{\widetilde\times}\, #2 \right)^{#3} }}
\newcommand{\vev}[1]{\ensuremath{\left< #1 \right> }}
\newcommand{\F}{\ensuremath{\mathcal{FT}}}
\renewcommand{\mod}[1]{\ensuremath{\left| #1 \right| }}
\newcommand{\var}[2][]{\ensuremath{\frac{\delta #1}{\delta #2}}}
\newcommand{\tr}[2][]{\ensuremath{\mathrm{tr}^{#1}\!\left\{ #2 \right\}}}

\newcommand{\se}{\ensuremath{s_\varepsilon}}
\newcommand{\seb}{\ensuremath{\bar s_\varepsilon}}

\newcommand{\sa}{\ensuremath{s_\alpha}}
\newcommand{\sab}{\ensuremath{\bar s_\alpha}}
\newcommand{\pp}{\ensuremath{\phi_+}}
\renewcommand{\pm}{\ensuremath{\phi_-}}
\newcommand{\uone}{\ensuremath{{U(1)}}}

\newcommand{\vph}{\varphi}
\newcommand{\Op}{\ensuremath{\mathcal{O}}}
\newcommand{\W}{\ensuremath{\mathfrak{W}}}
\renewcommand{\P}{\ensuremath{\mathfrak{P}}}

\newcommand{\ket}[1]{\ensuremath{\vert #1 \rangle}}

\newcommand{\be}{\begin{equation}}
\newcommand{\ee}{\end{equation}}
\renewcommand{\eqref}[1]{Eq.~(\ref{#1})}
\newcommand{\eqsref}[1]{Eqs.~(\ref{#1})}

\renewcommand{\L}{\ensuremath{\mathcal L}}
\newcommand{\LGF}{\ensuremath{\mathcal L_{GF}}}
\newcommand{\LA}{\ensuremath{\L_A}}
\newcommand{\LM}{\ensuremath{\L_M}}
\newcommand{\LYM}{\ensuremath{\L_{YM}}}

\renewcommand{\S}{\ensuremath{\mathcal{S}}}

\newcommand{\Smag}{\ensuremath{\mathcal{S}_{MAG}}}

\newcommand{\D}{\ensuremath{\mathsf{D}}}
\renewcommand{\O}{\ensuremath{\mathcal{O}}}
\newcommand{\N}{\ensuremath{\mathcal{N}}}
\newcommand{\G}{\ensuremath{\mathcal{G}}}

\DeclareMathOperator{\Ker}{Ker}
\DeclareMathOperator{\Ima}{Im}

\begin{document}

\title{Infrared Saturation and Phases of Gauge Theories with BRST Symmetry}

\author{Valentin Mader}
\email{valentin.mader@uni-graz.at}
\affiliation{Institut f\"ur Physik, Karl-Franzens-Universit\"at Graz, A-8010 Graz, Austria}

\author{Martin Schaden}
\email{mschaden@rutgers.edu}
\affiliation{Department of Physics, Rutgers, The State University of New Jersey, 
101 Warren Street, Newark, New Jersey - 07102, USA }

\author{Daniel Zwanziger}
\email{dz2@nyu.edu}
\affiliation{Physics Department, New York University, 4 Washington Place, New York, NY 10003, USA}

\author{Reinhard Alkofer}
\email{reinhard.alkofer@uni-graz.at}
\affiliation{Institut f\"ur Physik, Karl-Franzens-Universit\"at Graz, A-8010 Graz, Austria}

\begin{abstract} 

\noindent{\bf Abstract:} 
We investigate the infrared limit of the quantum equation of motion of the gauge boson propagator in various gauges and models with a BRST symmetry. We find that the saturation of this equation at low momenta distinguishes between the Coulomb, Higgs and confining phase of the gauge theory. The Coulomb phase is characterized by a massless gauge boson. Physical states contribute to the saturation of the transverse equation of motion of the gauge boson at low momenta in the Higgs phase, while the saturation is entirely due to unphysical degrees of freedom in the confining phase. This corollary to the Kugo-Ojima confinement criterion in linear covariant gauges also is sufficient for confinement in general covariant gauges with BRST- and anti-BRST symmetry, maximal Abelian gauges with an equivariant BRST symmetry, non-covariant Coulomb gauge and in the Gribov-Zwanziger theory.

\end{abstract}
\maketitle

\section{Introduction}       
    
In their seminal work, \cite{Kugo:1979gm}, Kugo and Ojima develop the covariant operator formalism for gauge theories in linear covariant gauge. On the assumption of an unbroken BRST symmetry, they construct the physical Hilbert space of the theory and formulate a criterion for color-confinement. The Hilbert space of a gauge theory is defined by the cohomology of $s$, the nil-potent generator of  BRST transformations, 
    \be \mathcal H_{phys} = \overline{\Ker{s}/\Ima{s}}\,. \ee
The confining phase of a gauge theory according to \cite{Kugo:1979gm} is characterized by an unbroken global color symmetry  \emph{and} the absence of massless gauge bosons.  In contrast to criteria based on the behavior of the Wilson loop, the Kugo-Ojima confinement criterion does not depend on the matter content of the theory.

Kugo and Ojima consider the conserved color current operator $J_\mu^a$ of Yang-Mills theory in linear covariant gauge,
\be J_\mu^a = -\partial_\nu F_{\nu\mu}^a + i s (D_\mu\bar c)^a \,.
\label{1_KO:colorcurrent} \ee
$J_\mu^a$ is related to the canonical Noether current $j_\mu^a$ by the quantum equation of motion (QEoM) of the gluon, 
\be\label{decomj}
j_\mu^a=J_\mu^a -\frac{\delta S}{\delta A^a_\mu}\ ,
\ee
where $S$ is the gauge-fixed action.
According to Kugo and Ojima, color confinement is realized if neither of the two currents
\be \G_\mu^a = -\partial_\nu F_{\nu\mu}^a\qquad \text{and}\qquad 
\N_\mu^a =  i s (D_\mu\bar c)^a  \label{1_KO:currents}\ee
create massless excitations. The corresponding charge operators
\be  \mathrm{G}^a = \int\!d^3x\,\G_0^a \qquad \text{and}\qquad  
\mathrm{N}^a = \int\!d^3x\,\N_0^a \label{1_KO:charges}\ee
are then both well-defined. The global color charge 
$Q^a =  \mathrm{G}^a + \mathrm{N}^a $ then also is well-defined and 
vanishes on the physical Hilbert space. 

Following \cite{Kugo:1979gm,Kugo:1995km}, we introduce the function $u(p^2)$ by the correlator 
\be  i\vev{(D_\mu c)^a\,(D_\nu \bar c)^b}_\F = 
\delta^{ab}\left(T_{\mu\nu}\,u(p^2) - L_{\mu\nu} \right)\label{KOcorr1} \,,\ee
where  $L_{\mu\nu} = p_\mu p_\nu/p^2$ and  $T_{\mu\nu} = \delta_{\mu\nu} - L_{\mu\nu}$ 
are longitudinal and transverse projectors, and $\F$ means Fourier transform.  The BRST exact charge 
$\mathrm N^a$ is well-defined only if,
\be u(p^2) \xrightarrow{p^2\rightarrow 0} -1\,. \label{1_KO:cond1} 
\ee

Together with the absence of massless vector bosons, \eqref{1_KO:cond1} is a \emph{sufficient} condition for color confinement\cite{Kugo:1995km}.
Here we express the confinement criterion of  Kugo and Ojima in terms of the saturation of the gluonic QEoM, 
\be
\vev{ A_\mu^a(x) {\delta S \over \delta A_\nu^b(y)}} = \delta_{\mu \nu} \delta^{ab} \delta(x-y),
\ee
at vanishing momentum. 
With the classically conserved current of global color symmetry $j_\mu^a$ of \eqref{decomj}, this equation in linear covariant gauge reads,
\begin{multline} \delta^{ab}\delta_{\sigma\mu} = -\vev{A_\sigma^a\,\ \partial_\nu F_{\nu\mu}^b}_\F  -\vev{A_\sigma^a\,(j_\mu^b-is(D_\mu\bar c)^b)}_\F \,.\\ \label{1:QEoM}\end{multline}
If $\G_\mu^a$ does not create a massless vector boson,  the first correlation function in \eqref{1:QEoM} vanishes in the infrared limit $p^2\rightarrow 0$ and the current,
\be\label{currtilde}
\tilde j_\mu^b=j_\mu^b-is(D_\mu\bar c)^b\ ,
\ee
must saturate \eqref{1:QEoM} at vanishing momentum. The current $\tilde j_\mu^b$ is physically equivalent to the classical color current $j_\mu^b$ because they differ by a BRST-exact term only. Condition (\ref{1_KO:cond1})  for the correlator (\ref{KOcorr1}) implies that 
\be 
i\vev{A_\sigma^a\,\ s(D_\mu\bar c)^b}_\F \xrightarrow{p^2\rightarrow 0}  \delta^{ab}\delta_{\mu\nu} \,.\ee
Thus: \emph{If the Kugo-Ojima criterion is fulfilled, the gluonic QEoM \eqref{1:QEoM} in linear covariant gauge at vanishing momentum is saturated by BRST-exact states.} Since the Physical Hilbert space does not include BRST-exact states, only unphysical states contribute to the saturation of the gluonic QEoM at vanishing momentum.
 
We therefore have the following \newline 
\emph{Proposition: 
In the confining phase of a gauge theory, unphysical states created by the color current $\tilde j_\mu^a$ saturate the gluonic QEoM at vanishing momentum.}

Since confinement allows only color singlet asymptotic states and any asymptotic state that contributes to the gluonic QEoM at vanishing momentum is in the adjoint color representation, the proposition clearly holds. An adjoint multiplet of physical asymptotic states on the other hand exists in Higgs and Coulomb phases. 
 Thus one can discriminate between the Higgs and confinement phase in linear covariant gauge by whether or not physical states contribute to the matrix element of the (generalized) color current $\tilde j_\mu^b$ at vanishing momentum.   

It is another matter to prove that the gluonic QEoM indeed \emph{is} saturated by unphysical states at vanishing momentum. Although we do not show this, we find that BRST-exact states in principle can saturate the gluonic QEoM at vanishing momentum in various gauges with a BRST or equivariant BRST symmetry.  In particular, we verify in these gauges that,
\begin{itemize}
\item[i)] if the theory does not describe the Coulomb phase, the gluonic QEoM
at vanishing momentum is saturated by the matrix element with a current that is physically equivalent to the conserved color current, and
\item[ii)] in non-Abelian gauge theories BRST-exact terms exist that can saturate the gluonic QEoM at low momentum.  
\end{itemize}

Since physical states contribute to the gluonic QEoM at vanishing momentum in Higgs and Coulomb phases, we conclude that saturation of the gluonic QEoM at vanishing momentum  by a BRST-exact term of the generalized current is a sufficient condition for confinement, provided that the BRST (or equivariant BRST) symmetry is unbroken.

In this article we identify the generalized current and the unphysical BRST-exact term 
in the gluonic QEoM in various gauges as well as in the Gribov-Zwanziger theory. The emergence of a similar pattern in all these models supports our proposition.

In a lattice theory with fundamental scalars, the Higgs and confining phases at finite lattice coupling 
$\beta$ are analytically connected  \cite{Fradkin:1978dv,Lang:1981qg,Caudy:2007sf}.  The situation is akin to a vapor-liquid transition below the critical point, and a (first-order)  transition does seem to occur at sufficiently large $\beta$. It is therefore not clear whether the two phases remain analytically connected in the continuum limit.  Since color charge is screened by a Higgs (or quark-) field in the fundamental representation, the asymptotic behavior of the Wilson loop is expected to always be perimeter-like and cannot be used to distinguish between phases. As the liquid-vapor transition below the critical point demonstrates, the absence of an order parameter does not necessarily imply that the free energy is analytic everywhere. Gauge-dependent criteria for Higgs and confining phases give different critical curves \emph{below} the critical $\beta$-value, but are remarkably consistent \emph{above} this critical point \cite{Caudy:2007sf}.  Analogous  results have been obtained in a semi-classical 
continuum analysis of a  non-Abelian Higgs model within the Gribov-horizon \cite{Capri:2012ah,Capri:2013oja}.     

However, to unambiguously distinguish between a Higgs and a confining phase, we in this article consider only Yang-Mills theory without fundamental matter. This gauge theory is either  in a Coulomb, a Higgs  or a confining phase and the asymptotic behavior of the Wilson- (or of the t'Hooft-) loop \cite{'tHooft:1979bi}  distinguishes between the latter two. The investigation of transitions between these phases is beyond the scope of the present article.

We will consider Yang-Mills theory in  linear covariant (LCG), generalized linear covariant (GLCG), maximal Abelian (MAG), Coulomb (CG) and minimal Landau (GZ) gauge. Some  known results for these gauges are summarized below.

The best investigated  Linear Covariant Gauge (LCG) is  Landau gauge. Another corollary 
of the Kugo-Ojima confinement criterion \cite{Kugo:1995km} in this gauge relates the
infrared limit of the ghost  dressing function $G(p^2)$ to $u(p^2)$,  $\lim_{p^2\to
0}G(p^2)^{-1} = 1+ \lim_{p^2\to 0} u(p^2)$.  When \eqref{1_KO:cond1} is fulfilled,  the ghost dressing function diverges in the  infrared, and the ghost propagator at vanishing momentum is more singular than a  massless pole. An exact infrared analysis of the whole tower of Dyson-Schwinger (DSE) and of Exact Renormalization Group Equations (ERGE) confirms the existence of solutions with this infrared behavior of the ghost propagator and a corresponding infrared suppressed gluon propagator, see for instance
refs.\  \cite{vonSmekal:1997is,Atkinson:1997tu,Lerche:2002ep,Zwanziger:2002xc,Fischer:2002hna,
Pawlowski:2003hq,Alkofer:2004it,Huber:2007kc,Fischer:2008uz,Huber:2012kd}.  The solutions for which the Kugo-Ojima criterion is fulfilled have a
power-like enhancement of the ghost propagator with related infrared exponents
of  other Green functions determined by an infinite tower of scaling relations
\cite{Alkofer:2004it,Huber:2007kc}.  According to numerical studies of Yang-Mills theory in 2d \cite{Maas:2007uv,Cucchieri:2007rg,Cucchieri:2011um}, the ghost and gluon propagators exhibit this scaling behavior, and moreover a strict analytic bound \cite{Zwanziger:2012xg} implies that in $2d$ the gluon propagator vanishes at zero momentum $D(0) = 0$.

However in $3d$ and $4d$ another one-parameter family of solutions to the DSEs and ERGEs also exists that is best parameterized by the
value of $G(0)^{-1}\neq 0$. The ghost propagator of these solutions is only quantitatively enhanced and the gluon propagator is infrared finite 
\cite{Aguilar:2008xm,Fischer:2008uz,Boucaud:2011ug,Huber:2012kd}.  Lattice gauge theory studies in three and four dimensions observe only this  infrared finite behavior
\cite{Sternbeck:2005tk,Bogolubsky:2009dc,Maas:2011se,Oliveira:2012eh} of  lattice 
propagators. But a whole one-parameter family of solutions can also be obtained on the lattice by tuning the lowest eigenvalue of the Faddeev-Popov operator \cite{Sternbeck:2012mf} of
the gauge fixing.  The origin of this multitude of solutions is unresolved and it has been suggested\cite{Maas:2011se} that the value of $G(0)^{-1}$ could be considered an additional gauge parameter. Numerical solutions over the whole momentum range are
only available for truncated DSEs and ERGEs. 

Within the error of the employed approximation and/or truncation, both types of
solutions lead to very similar  phenomenology \cite{Blank:2010pa} and confine static
quarks \cite{Fister:2013bh}. Unbroken BRST symmetry is essential for the Kugo-Ojima
confinement criterion.  Without recourse to a preserved nil-potent symmetry it is
difficult to identify the unphysical sector of such truncated models. However, our proposition that confinement is a Higgs mechanism in the unphysical sector of the theory can be formulated in the absence of a nil-potent symmetry and may hold in all these scenarios.

In the minimal Coulomb gauge, the dressing function of the ghost propagator of Yang-Mills
theory is more divergent than a simple massless pole, effectively leading to a confining
color-Coulomb potential \cite{Zwanziger:2002sh}. The infrared divergence of the
instantaneous ghost propagator in this gauge has been verified by other calculations in
the continuum \cite{Szczepaniak:2001rg,Epple:2006hv,Schleifenbaum:2006bq} and on the
lattice \cite{Burgio:2012bk}. For a thorough discussion of the current status see
\cite{Watson:2011kv}. 

The dual superconductor picture of the QCD vacuum \cite{Mandelstam:1974pi,'tHooft:1995fi}
is the motivation for considering Yang-Mills in maximal Abelian gauge (MAG),
\cite{Kronfeld:1987ri,Kronfeld:1987vd,Schaden:1998hz,Schaden:1999ew,Shinohara:2001cw}.
This gauge discriminates between the Cartan subalgebra and the coset space of the gauge
group and the partial gauge fixing  breaks the local $SU(N)$ invariance down to the Cartan
subgroup $\uone^{N-1}$.  The hypothesis of Abelian dominance \cite{Ezawa:1982bf} states
that the Cartan gluons dominate long-range interactions. This has been observed in lattice
simulations \cite{Gongyo:2013sha,Gongyo:2012jb} and is also corroborated by an infrared
analysis of the  functional equations \cite{Huber:2009wh,Alkofer:2011di}. Furthermore, the
Cartan gluons also dominate at large momenta  \cite{Quandt:1997rw, Gracey:2005vu}.
A detailed understanding of the relation between the infrared behavior of
Green's functions and confinement nevertheless is lacking in the MAG. It is of
interest that a renormalization group analysis of interpolating gauges found that Abelian
gauges form an invariant subspace that is not analytically connected to linear
covariant gauges \cite{Hata:1992np}. This explains why a literal
interpretation of the Kugo-Ojima confinement criterion fails for this class of gauges
\cite{Suzuki:1983cg}. Our proposition that unphysical states saturate the gluonic  QEoM at vanishing momentum can nevertheless be realized. Contrary to Abelian gauge theories, the Abelian current of non-Abelian gauge theories includes an operator that only creates unphysical states that could saturate the gluonic QEoM.  

The Gribov-Zwanziger framework
\cite{Gribov:1977wm,Zwanziger:1988jt,Zwanziger:1989mf,Zwanziger:1992qr,Vandersickel:2012tz}
restricts the path-integral to the first Gribov region of LCG and Coulomb gauge. Even
though it drastically changes the gauge-fixed action and breaks BRST symmetry spontaneously,
\cite{Maggiore:1993wq,Dudal:2012sb}, it does not change the form of
the DSEs in Landau gauge \cite{Zwanziger:2002sh}, and the
infrared exponent of the scaling solution is that of ordinary Landau gauge\cite{Huber:2009tx}. We find
that the color current of this model includes a BRST-exact contribution.
The latter in fact saturates the gluonic QEoM at vanishing momentum already at tree level. However,
it is difficult to verify that this BRST-exact operator only creates unphysical states
since the BRST-symmetry of this model is spontaneously broken.

Yang-Mills theory is expected to confine in all these gauges, and one hopes that the
underlying mechanism can be characterized in a gauge-invariant fashion. Although the
dynamics may be different, our proposition --- concerning saturation of the gluonic QEoM at
vanishing momentum by unphysical states --- can be realized in all of them. Other
similarities include that the dielectric function of the QCD vacuum appears to be related
to the divergent dressing function of the ghost propagator in Coulomb gauge
\cite{Reinhardt:2008ek}, and that in Landau gauge the Kugo-Ojima criterion is related to
the Gribov-Zwanziger scenario, \cite{Dudal:2009xh}.

The present article is organized as follows:  In Sec.~\ref{sec_qed}, we examine the 
gluonic QEoM in
Abelian gauge theory in linear covariant gauges, and review the arguments
that  distinguish between Coulomb and Higgs  phases in this specific case. In
Sec.~\ref{sec_acoulomb}, we examine the Abelian Coulomb phase in greater detail, and in
Sec.~\ref{sec_higgs}, we explicitly verify the implications of a spontaneously broken \uone-symmetry in
the 't Hooft gauge. In Sec.~\ref{sec_ko} the Kugo-Ojima confinement criterion for LCG is
reviewed. We then proceed to generalize and adapt this criterion to other gauges: to generalized covariant gauges in Sec.~\ref{sec_cc}, to covariant but non-linear MAG in
Sec.~\ref{sec_mag}, and to the non-covariant Coulomb gauge in Sec.~\ref{sec_CG}. In all these
gauges a Kugo-Ojima-like confinement  criterion is formulated. In Sec.~\ref{sec_GZ} we examine  signatures of
confinement in the Gribov-Zwanziger (GZ) theory. Sec.~\ref{sec_concl} summarizes our results. Conventions and some  technical details are deferred to three appendices.

\section{Abelian Gauge Theories\label{sec_qed}}
In Abelian gauge theories one can, of course, dispense with a BRST construction
of observables \cite{Gupta:1949rh,Bleuler:1950cy}. However, identifying the physical sector by
a BRST symmetry is readily extended to non-Abelian gauge theories, and to non-canonical
quantization.

To this end we consider Abelian gauge theories in general linear covariant gauges,
\begin{align}\label{L_QED}
 \L_\uone&= \LA  + \LM +s\left(\bar c\left(\frac \xi 2 b  -i\partial_\mu A_\mu +i \gamma(\phi,\dots)\right)\right)\nonumber\\
&=\LA  + \LM +\frac \xi 2 b^2  -ib\partial_\mu A_\mu+i b\gamma(\phi,\dots) \\
& \hspace{30mm}+i\bar c\partial^2 c-i\bar c s\gamma(\phi,\dots)\, .\nonumber\end{align}
Here $\xi$ is a gauge parameter and $b,c,$ and $\bar c$ are the Nakanishi-Lautrup (NL) and
(anti-commuting) ghost and anti-ghost fields. The local function $\gamma(\dots)$ of canonical
dimension $2$ and vanishing ghost number is a  polynomial of bosonic matter fields $\phi$ that
does not depend on the gauge connection $A_\mu$ or the NL field $b$. The matter part, \LM, is
invariant under \uone-gauge transformations and may include covariantly coupled charged
fermions and bosons.  

The variation $s$ generates the nilpotent BRST symmetry \cite{Becchi:1975nq,Tyutin:1975qk}
of $\L_\uone$ (\ref{L_QED}), 
\begin{align}\label{qedquartet}
s A_\mu &=\partial_\mu c \, , & s c&=0\, ,& 
s \bar c &=b \, , & s b &=0\, .
\end{align}
Under the BRST variation  $s$, charged matter fields vary by an infinitesimal \uone-gauge
transformation with the ghost field $c(x)$ as variation. The longitudinal gauge field, ghost
$c$, anti-ghost $\bar c$ and NL field $b$ form the elementary
BRST quartet\cite{Kugo:1979gm,Peskin:1995ev}. We note that one can define an anti-BRST
variation in this Abelian setting by
\begin{align}\label{qedaquartet}
\bar s A_\mu &=\partial_\mu \bar c \, , & \bar s \bar c&=0 \, ,&
\bar s c &=- b \, , & \bar s b &=0\, ,
\end{align}
with an obvious extension to matter fields.  The generators of BRST and of anti-BRST 
transformations are nilpotent and anticommute,  
$s^2=\bar s^2=\acom{s}{\bar s}=s \bar s+\bar s s=0$. 
The conserved BRST and anti-BRST charges corresponding to the transformations in
\eqref{qedquartet} and \eqref{qedaquartet} in the Abelian case may be represented by the
functional derivative operators,
\begin{subequations}\label{BRSTAbelian}
\begin{align}\label{QQAbelian}
Q_{BRST}&=\int d^4x\Biggl(-c(x)\partial_\mu \frac{\delta}{\delta A_\mu(x)}+
b(x)\frac{\delta}{\delta \bar c(x)} \\
	& \hspace{30mm}+\sum_ M(s \phi_M(x)) \frac{\delta}{\delta \phi_M(x)}
	\Biggr)\nonumber \ ,\\
 \bar Q_{BRST}&=\int d^4x\Biggl(-\bar c(x)\partial_\mu \frac{\delta}{\delta A_\mu(x)}-
 b(x)\frac{\delta}{\delta c(x)}\\
	  & \hspace{30mm}+\sum_ M(\bar s \phi_M(x)) \frac{\delta}{\delta \phi_M(x)}
	  \Biggr)\ , \nonumber
\end{align}
\end{subequations}
where the sums run over all matter fields $\phi_M(x)$. The ghost number,
\be\label{ghostN} 
\Pi=\int d^4x\left( c(x)\frac{\delta}{\delta c(x)}-
\bar c(x)\frac{\delta}{\delta \bar c(x)}\right)\, ,
\ee
also is conserved.

The nilpotent BRST symmetry allows one to define the subset $\P$ of physical operators by the 
cohomology~\cite{Becchi:1975nq},   
\begin{align}\label{physOP}
\P&=\{\text{physical operators}\}\\ 
  &=\{\O;\  [Q_{BRST},\O]=0\ \text{and}\ [\O,\Pi]=0\} \nonumber \\
  & \hspace{10mm} / \{[Q_{BRST},\O];\  [\O,\Pi]=\O\}\ . \nonumber
\end{align}   
Using a canonical construction, it was shown \cite{Kugo:1979gm} that negative norm states
associated with asymptotic BRST quartets are unphysical. The elementary quartet thus is not
observable. Transversely polarized photons, on the other hand, are physical. Instead of
constructing the physical asymptotic Hilbert space directly, we prefer to define the space of
physical operators, which is a slightly more flexible point of view that can be extended to
spaces other than four-dimensional Minkowski spacetime.  With a BRST invariant ground state,
the two approaches are equivalent in Minkowski space. 

Since matter transforms covariantly under \uone-gauge transformations, $A_\mu\rightarrow A_\mu
+\partial_\mu\theta$, and $\LM$ is an invariant, the conserved global \uone-current,
$j_\mu^\uone$, is obtained from the matter part of the action alone,
\be
{j_\mu}^{\hspace{-.4em}\uone}(x)=-\frac{\delta \S_M}{\delta A_\mu(x)}\, ,
\label{U1-current}
\ee
with $\S_M=\int d^4x \LM$. 

The QEoM of the Abelian gauge boson propagator in linear covariant gauges
reads,
\begin{align}
 \delta_{\sigma\mu}\delta(x-y) & = \vev{A_\sigma(y)\,\var[S_\uone]{A_\mu(x)}} \nonumber \\
    & = -\vev{ A_\sigma(y)\,  \partial_\nu F_{\nu\mu}(x)} -\vev{A_\sigma(y)\,
    {j_\mu}^{\hspace{-.4em}\uone}(x)} \nonumber\\
    & \qquad + i\vev{ A_\sigma(y)\, s \partial_\mu \bar c(x)}  \, . \label{qed_dse}
\end{align}
 It depends on the classically conserved current ${j_\mu}^{\hspace{-.4em}\uone}(x)$ of
\eqref{U1-current} and holds for renormalized as well as for bare fields. The last,
longitudinal, term on the rhs in \eqref{qed_dse} arises from the linear covariant gauge fixing
in \eqref{L_QED} and is BRST-exact. It has no physically observable effects and may be included in the definition of the current ${\tilde j_\mu}^\uone(x)={j_\mu}^{\hspace{-.4em}\uone}(x)-i s \partial_\mu \bar
c$. In close analogy to the non-Abelian case discussed below one can use \eqref{qedaquartet} to write, $s \partial_\mu \bar
c = s \bar s A_\mu$. 

The first term on the rhs of \eqref{qed_dse} is transverse due to the antisymmetry of the 
field strength tensor. Fourier transformation (denoted by $\vev{...}_\F$) of \eqref{qed_dse} 
and longitudinal  ($L_{\mu\nu} = {p_\mu p_\nu}/{p^2}$) and transverse 
($T_{\mu\nu} = \delta_{\mu\nu} - L_{\mu\nu}$) projection give the identities,
\begin{subequations}\label{decompose}
\begin{align}
\label{longitudinal}
L_{\sigma\mu}&=\vev{A_\sigma(y)\, i \partial_\mu s\bar c(x)}_\F- 
L_{\mu\rho}\vev{A_\sigma(y)\, {j_\rho}^{\hspace{-.4em}\uone}(x)}_\F \, ,
\\
\label{transverse}
 T_{\sigma\mu}  &= -\vev{ A_\sigma(y)\, \partial_\nu F_{\nu\mu} }_\F 
 -T_{\mu\rho}\vev{ A_\sigma(y)\,{j_\rho}^{\hspace{-.4em}\uone}(x)}_\F\, .
\end{align}
\end{subequations}

Using the equation of motion of the NL field, \eqref{longitudinal} yields the Ward identity for
the longitudinal photon propagator,
\begin{multline}
\xi\, p_\sigma= p^2 p_\nu \,\vev{A_\sigma(y)\,A_\nu(x)}_\F -ip^2\vev{A_\sigma(y)\,
\gamma(x)}_\F \\ 
	- i\xi\vev{A_\sigma(y)\,\partial_\nu {j_\nu}^{\hspace{-.4em}\uone}(x)}_\F \,, 
	\label{qed_wi} 
\end{multline}
where $\gamma(x)=\gamma(\phi(x),\dots)$ is the local function of the fields in the BRST exact
term of \eqref{L_QED}.

The correlator of the gauge field with the divergence of the field strength tensor is
transverse and is described by a Lorentz invariant function $f(p^2)$,
\be\label{transdef} 
T_{\sigma\mu} f(p^2):=-\vev{ A_\sigma(y)\,  \partial_\nu F_{\nu\mu}(x)}_\F\ ,
\ee
which for $p^2\rightarrow 0$ determines the phase of the model. $f(0)\neq 0$ implies a pole 
at $p^2=0$ due to a massless transverse vector boson in the correlator
\be\label{masslessvb}
\vev{ A_\sigma(y)\,  F_{\nu\mu}(x)}_\F=-i (\delta_{\sigma\mu} p_\nu-
\delta_{\sigma\nu}p_\mu) \frac{f(p^2)}{p^2}\ .
\ee
A model with $f(0)\neq 0$ thus has a massless photon and describes a Coulomb phase.  The Abelian nature of the field strength is not essential for inferring a massless pole in 
\eqref{masslessvb}. One merely exploits the Poincar{\'e} invariance and the 
antisymmetry of the curvature $F_{\mu\nu}$.

In the Abelian case, $f(p^2)$ determines the transverse part of the vector boson propagator,
\be\label{photontransverse}
T_{\mu\nu} \, \vev{A_\sigma(y)\,A_\nu(x)}_\F=\frac{f(p^2)}{p^2} T_{\sigma\mu}\    ,
\ee
and the photon is massless only if $f(0)>0$. Insertion of the definition, \eqref{transdef}, into
\eqref{transverse} shows that $f(p^2)$ also completely determines the \emph{transverse} current
matrix element,
\be\label{current}
T_{\mu\nu}\vev{ A_\sigma(y)\,{j_\nu}^{\hspace{-.4em}\uone}(x)}_\F=(f(p^2)-1)T_{\sigma\mu}\ .
\ee
 If the current saturates \eqref{transverse} in the infrared,
\be\label{higgscond}
T_{\mu\nu}\vev{ A_\sigma(y)\,{j_\nu}^{\hspace{-.4em}\uone}(x)}_\F\stackrel{p^2\rightarrow 0}
{\xrightarrow{\hspace*{1cm}}}-T_{\sigma\mu}\ ,
\ee
the vector boson is short-ranged, and $f(0)=0$.
 
These relations hold for any Abelian gauge theory in linear covariant gauges. $f(0)\neq 0$
characterizes a model describing a Coulomb phase with a massless photon. We next examine Abelian
gauge theories in the Coulomb and Higgs phase in more detail. In Sec.~\ref{sec_mag} we consider
an Abelian gauge theory that confines color charge and investigate possible signatures of this
phase.

\subsection{The Coulomb phase\label{sec_acoulomb}}
The Coulomb phase is characterized by an unbroken Abelian gauge symmetry \emph{and} the failure
of \eqref{higgscond}, {\it i.e.}, failure of the current contribution to saturate 
\eqref{transverse} in the infrared. Since the Abelian gauge symmetry is unbroken,  the
correlation function  \vev{ A_\sigma(y)\,{j_\nu}^{\hspace{-.4em}\uone}(x)} is transverse in any
covariant gauge. \eqref{longitudinal} and the Ward identity \eqref{qed_wi} have the form
\begin{multline}\label{longCoulomb}
 L_{\sigma\mu}=p_\mu\vev{A_\sigma(y)\, b(x)}_\F \\ \text{and}\ \ \xi\, 
 \frac{p_\sigma}{p^2}= p_\nu \,\vev{A_\sigma(y)\,A_\nu(x)}_\F\ .
\end{multline}
 The elementary quartet is free and massless:
\be\label{propquartet}
\vev{ A_\mu(y)\,b(x)}_\F=\vev{\partial_\mu c(y)\, \bar c(x)}_\F=\frac{p_\mu}{p^2}  \,.
\ee
From \eqref{photontransverse} and \eqref{longCoulomb} the photon propagator is given by
\be \label{propCoul}
\vev{A_\mu(y)\,A_\nu(x)}_\F = \frac{f(p^2)}{p^2} T_{\mu\nu}+ \frac{\xi }{p^2} L_{\mu\nu}\ . 
\ee
The photon is massless with $f(0)>0$, because \eqref{higgscond} does not hold.
 
It is interesting to note that in the canonical formalism $f(0)\neq 0$ implies that the
electromagnetic charge operator,
\be\label{qedQ}
Q=\int d^3x {j_0}^{\hspace{-.4em}\uone}(x)\, ,
\ee
is not well defined. Up to terms proportional to the photon equation of motion, this charge 
is equivalent to
\begin{multline}\label{Qfalse}
Q\equiv\tilde Q=\int d^3x (i\partial_0 b(x)-\partial_\nu F_{\nu 0}(x)) \\ =
\int d^3x i\partial_0 b(x)+\int_{S_\infty} d{\mathbf \sigma}_i F_{0i}=\N+\mathcal G\ .
\end{multline}
Due to the antisymmetry of the field strength tensor, $F_{\nu\mu}=-F_{\mu\nu}$, the
current $-\partial_\nu F_{\nu \mu}(x)$ and corresponding charge $\mathcal G$ are themselves 
conserved. Furthermore, the equal time commutator of $\mathcal G$ with any \emph{local} physical operator 
$\Phi(x)\in\P$ vanishes,
\be\label{causloc}
[\Phi(x),\mathcal G]=0\  \text{ for all \emph{local}}\  \Phi(x)\in\P\ ,
\ee
because causalility requires operators with spatial separation to commute.
One thus has that,
\begin{align}\label{chargecomm}
\com{Q}{\Phi(x)}&\equiv \com{\N+\mathcal G}{\Phi(x)}=\com{\acom{\mathcal C}{Q_{BRST}}}
{\Phi(x)}\\
& =\acom{\mathcal C}{\com{Q_{BRST}}{\Phi(x)}}+\acom{Q_{BRST}}{\com{\mathcal C}{\Phi(x)}}
\nonumber\\
&=\acom{Q_{BRST}}{\com{\mathcal C}{\Phi(x)}} \nonumber
\end{align}
 for any \emph{local} $\Phi(x)\in\P$.  Here
 $\mathcal C=i\int d^3 x \partial_0 \bar c =\int d^3 x \pi_c(x)$ 
 is the canonical conjugate of the ghost operator at vanishing momentum and 
 $\N = \acom{Q_{BRST}}{{\mathcal C}}$. 
In deriving \eqref{chargecomm} one uses causality, the Jacobi identity and that $\N$ is 
BRST exact. All \emph{local physical} operators $\Phi(x)\in \P$ thus are uncharged and 
physical operators creating charged particles like the electron necessarily are not local.  
(NB: To compare with non-abelian gauge theories, note that $Q_{BRST}$  and ${\mathcal C}$ may 
be  replaced by $\bar Q_{BRST}$ and   $\bar{\mathcal C}=\int d^3 x \pi_{\bar c}(x)$ in 
\eqref{chargecomm}.) One \emph{can} construct  non-local charged physical states in QED 
because infrared photon states are almost degenerate with the ground state 
\cite{Bloch:1937pw, Kulish:1970ut, Gervais:1980bz, Bagan:1999jf}. The massless vector boson of 
the Coulomb phase thus prevents one from concluding from \eqref{chargecomm} that all 
physical states are uncharged. 
    
\subsection{The Abelian Higgs phase \label{sec_higgs}}
A ``spontaneously broken'' Abelian gauge theory in the Higgs phase satisfies 
\eqref{qed_dse} differently. From the general discussion one expects that  $f(0)=0$, the vector
boson is massive and the current saturates \eqref{transverse} at low momenta, {\it i.e.}, 
\eqref{higgscond} holds. In the Higgs phase one also expects
(unphysical) massless excitations. We explicitly verify this scenario in QED with charged scalar
matter whose self-interactions are described by a Higgs potential with quartic coupling
$\lambda$ and a negative quadratic term proportional to $-4\lambda v^2$,
\begin{subequations}
\label{LHiggs}
    \begin{align} 
\label{raw}
    \LM^\text{Higgs}& = \frac{1}{2} (D_\mu \Phi)^* (D_\mu \Phi) + \lambda \left( \mod{\Phi}^2   
    - v^2\right)^2  \\ 
	  &= \frac 1 2 \left( (\partial_\mu \pp)^2 + (\partial_\mu\pm)^2 \right) 
	  + \frac {m^2}{2}\,A_\mu^2 + m\, \phi_- \partial_\mu A_\mu \nonumber\\
		& + g \, A_\mu \left( \pm \partial_\mu \pp-\pp \partial_\mu \pm  \right) 
		+ gm \, A_\mu^2 \pp  \label{higgs_lag}\\
	& + \frac{g^2}{2}\,A_\mu^2 \left( \pp^2 + \pm^2 \right) + 
	\lambda \left( \pp^2 + \pm^2 + 2 \pp v \right)^2  \,.  \nonumber
    \end{align}
\end{subequations}
In the Higgs phase with $v>0$ we parameterize the fields by,  $\Phi = \phi + v ,\, \pp = {\frac
1 2} (\phi^*+\phi) ,\, \pm  ={\frac i 2}( \phi^*-\phi) $. The tree level photon mass is $m =
gv$ and that of the Higgs field $\pp$ is $m^2_H=8\lambda v^2$. $\pm$ is massless and couples to
the longitudinal photon. The model is invariant under local gauge transformations $\delta A_\mu
= \partial_\mu \theta,\, \delta\Phi = i g \theta \Phi ,\, \delta\pp = - g \theta \pm ,\, \delta
\pm  =  g \theta \left( \pp +  v \right)$. Replacing $\theta(x)$ by the anti-commuting
ghost field one arrives at the BRST variations
    \begin{subequations}
\label{higgsBRS}
\be \label{elemquartet}
	s A_\mu = \partial_\mu c ,\qquad sc = 0 ,\qquad s \bar c  = b ,\qquad sb  = 0,  
\ee
\be \label{matter}
	s\pp  = -gc \,\pm  \, ,\qquad s \pm  =  g c \left( \pp + v \right)  \, .
\ee
\end{subequations}
A convenient gauge that eliminates the  bilinear coupling of $\pm$ to the longitudinal photon
is given by the BRST exact linear covariant gauge fixing term  \cite{'tHooft:1971rn},
    \begin{align}\label{tHooftgf}
	\LGF^\text{'t~Hooft} & = s\left(\bar c \left( -i \partial_\mu A_\mu + \frac \xi 2 b 
	+ i\xi m \pm \right) \right)\\
		& = \frac \xi 2 b'^2 - i b' \partial_\mu A_\mu - m \, \pm \partial_\mu A_\mu 
		+ \frac{\xi m^2}{2}\,\pm^2\nonumber \\
		& \hspace{10mm}+ i \bar c \left( \partial^2 - gm\xi \pp - m^2 \xi \right) c  
		\,,\nonumber
    \end{align}
where in the second expression the NL field has been shifted: $b = b' -i m \pm$.   
The classical Lagrangian of the Abelian Higgs model in linear covariant 't~Hooft gauge is 
\be\label{LH}
\L^\text{Higgs} = \LA+\LM^\text{Higgs} + \LGF^\text{'t~Hooft}\ .
\ee
Note that the BRST exact term $\LGF^\text{'t~Hooft}$  of \eqref{LH} explicitly breaks not 
only local  but also global \uone-symmetry. 

The QEoM of the photon propagator is given by \eqref{qed_dse} with the gauge invariant and classically  
conserved matter current
\begin{align}
{j_\mu}^{\hspace{-.4em}\uone}&=  \delta \Phi \,\var[\LM^\text{Higgs}]{\partial_\mu \Phi} 
+ \delta \Phi^* \,\var[\LM^\text{Higgs}]{\partial_\mu \Phi^*} \nonumber\\
&= g \left( \pp \partial_\mu \pm-\pm \partial_\mu \pp  \right) + m \partial_\mu \pm 
\label{higgs_curr} \\
& \hspace{15mm} -  A_\mu \left( g^2( \pp^2 + \pm^2)+m^2 +2 mg \pp \right)\,. \nonumber
\end{align}
The current is BRST invariant, and its divergence is unphysical because  the global gauge
invariance of the model is broken by BRST exact terms only. To leading order in the loop 
expansion one has
\be\label{invlong} 
s {j_\mu}^{\hspace{-.4em}\uone}\approx s(m\partial_\mu\pm- m^2 A_\mu)=gm\partial_\mu(c\pp)
\approx 0\ .
\ee 
In fact, the divergence $\partial_\mu{j_\mu}^{\hspace{-.4em}\uone}$  is BRST exact up to 
equations of motion. In leading approximation
\begin{multline}\label{divj}
\partial_\mu{j_\mu}^{\hspace{-.4em}\uone}\approx m\partial^2\pm-m^2 \partial_\mu A_\mu \\ 
\equiv m^2(\xi m\pm - \partial_\mu A_\mu) \equiv i m^2\xi b=i m^2\xi s\bar c\ ,
\end{multline}
 where the tree level QEoM of $\pm$ and of the NL field $b$ was used  to 
 obtain the intermediate expressions.   

In the broken phase,  the current contribution to \eqref{longitudinal} does not
vanish and in fact saturates it at low momenta.  This is the signature of a
``spontaneously broken'' gauge theory. Since the divergence of the current is BRST exact up to 
equations of motion, it does not create physically observable  Goldstone bosons and the 
 gauge theory is in a Higgs phase.  
In 't~Hooft gauges the mass of the unphysical scalar particle created by 
${j_\mu}^{\hspace{-.4em}\uone}$ 
depends on the gauge parameter $\xi$ and vanishes for $\xi=0$ only.

With $\gamma(x)=\xi m \pm(x)$, the Ward identity of \eqref{qed_wi} to leading order asserts,
\begin{align}
\label{longbroken}
\xi\, p_\sigma&= p^2 p_\nu \,\vev{A_\sigma(y)\,A_\nu(x)}_\F +i\xi m\vev{A_\sigma(y)\,
\partial^2\pm(x)}_\F \nonumber\\
	      & - i\xi\vev{A_\sigma(y)\,\partial_\nu {j_\nu}^{\hspace{-.4em}\uone}(x)}_\F
	      \nonumber\\
&\approx p^2 p_\nu \,\vev{A_\sigma(y)\,A_\nu(x)}_\F +i\xi m \vev{A_\sigma(y)\,
\partial^2\pm(x)}_\F \nonumber\\
    &-i\xi\vev{A_\sigma(y)\,\partial_\nu(m\partial_\nu\pm-m^2 A_\nu)(x)}_\F\nonumber\\
&= (p^2+\xi m^2)   p_\nu \,\vev{A_\sigma(y)\,A_\nu(x)}_\F\, , 
\end{align}
Note that  $\pm$ does not contribute to the Ward identity at tree level. This cancellation of 
mixing terms is a feature of 't~Hooft gauges.

However, the current matrix element in \eqref{longbroken} saturates \eqref{longitudinal} for
$p^2\rightarrow 0$ whereas it vanishes in the Coulomb phase in this limit. This is not a gauge
artifact and for $\xi\neq 0$ does not depend on the gauge parameter.  

 \eqref{longbroken} gives the tree-level longitudinal propagator in the Higgs phase:
\begin{multline}
   p_\nu \,\vev{A_\sigma(y)\,A_\nu(x)}_\F \approx \frac{\xi p_\sigma}{p^2 + \xi m^2} \\
      =\frac{p_\sigma}{m^2}-\frac{p^2p_\sigma}{m^2(p^2+\xi m^2)}\, , \label{higgs_wi}  
\end{multline}
which may be directly verified from the quadratic terms of the action \eqref{LH}. In the last
expression of \eqref{higgs_wi}, the $\xi$-independent term at $p^2=0$ arises from the current matrix element. The second term is the $\xi$-dependent negative-norm contribution
of  $\vev{A_\sigma(y)\, i\partial_\mu s \bar c(x)}\approx \frac{p^2}{p^2+\xi m^2}
L_{\sigma\mu}$. It is one of the correlators of the elementary BRST quartet and for 
$\xi m^2\neq0$ vanishes as $p^2\rightarrow 0$, leaving the current to saturate the 
\eqref{longitudinal} in the Higgs phase.  

In the Higgs phase the tree-level approximation to the function $f(p^2)$ defined by \eqref{transdef} is,
 \be 
f(p^2) \approx \frac{p^2}{p^2 + m^2} . \label{higgs_sat}
\ee
Since $f(0)=0$ the tranverse vector boson is short ranged in this phase, 
\be  T_{\mu\nu}  \vev{A_{\sigma}(y)\,A_\nu(x)}_\F \approx \frac{1}{p^2 + m^2}T_{\sigma\mu}  \,.
 \label{higgs_witrans}\ee
The current of \eqref{higgs_curr} thus saturates \eqref{transverse} at low momenta, for any value of the gauge parameter $\xi$ and
\eqref{higgscond} holds in the Higgs phase. Note that the physical Higgs particle and vector boson in this model are not charged.   

These examples illustrate (at tree level) the characteristics that distinguish the unbroken
Coulomb and ``spontaneously broken'' Higgs phases of Abelian gauge theories. If  the current
contribution saturates the transverse \emph{and} the longitudinal QEoM of the photon propagator at low
momenta, the model describes a ``spontaneously broken'' Higgs phase. If the current contribution
fails to saturate the transverse equation at low momenta, the Abelian gauge theory describes a Coulomb phase with a massless vector particle.  The (conserved) transverse part of the Abelian
current in our examples is BRST invariant and does not include BRST exact terms. At vanishing momentum it apparently creates only physical particles. This will change when we consider Abelian gauge theories  that confine color charge in Sec.~\ref{sec_mag}.  

First however, let us revisit non-Abelian gauge theories in Linear Covariant Gauges (LCG) for which Kugo and Ojima originally formulated their confinement criterion.

\section{Non-Abelian Gauge Theories}
\subsection {The Kugo-Ojima Confinement Criterion in Linear Covariant Gauges (LCG)\label{sec_ko}}
 While the photon is massless and atoms are readily ionized, gluons are only of short range and no hadron has been color-ionized. This confinement of color charge is one of the most prominent 
 features of unbroken non-Abelian gauge theories. One expects the confinement of color and the 
 absence of massless gluons to manifest themselves in solutions to the QEoM of the 
 vector boson propagator. We here give a short review of Kugo and Ojima's analysis 
 \cite{Kugo:1979gm} of  unbroken non-Abelian gauge theories in the linear covariant gauge (LCG). 

Yang-Mills theory in LCG is defined by the Lagrangian 
     \begin{align}
 \L_{LCG} &=  \LYM + s \left(\bar c^a\left( \frac{\xi}{2} b^a-i \partial_\mu A^a_\mu\right)\right)
 \nonumber\\
&= \LYM+\frac \xi 2 b^2 - i b^a \partial_\mu A_\mu^a  - i \partial_\mu\bar c^a (D_\mu c)^a \,.
\label{ko_lag}
	\end{align}
The nilpotent  BRST transformation in the non-Abelian case is given by
\begin{align}\label{ko_BRST}
      s A_\mu^a & = (D_\mu c)^a \, , & sc^a &  = -\frac{1}{2} \cp{c}{c}{a} \, , \nonumber \\ 
      s \bar c ^a & = b^a\, , & sb^a &= 0 \,, 
\end{align}
and is readily extended to covariantly coupled matter. The Lagrangian (\ref{ko_lag}) is also
invariant under an anti-BRST transformation generated by $\bar s$,  
\begin{align} \label{ko_aBRST}
      \bar s A_\mu^a & = (D_\mu \bar c)^a\,,  
      & \bar s\bar c^a  &= -\frac{1}{2} \cp{\bar c}{\bar c}{a}\,,\nonumber\\
      \bar s c ^a &  = - b^a - \cp{\bar c}{c}{a} \,, & \bar s b^a &= \cp{b}{\bar c}{a} \,.
\end{align}
It is in addition invariant under \emph{global} color rotations and 
preserves ghost number. The BRST and anti-BRST transformations are nilpotent and  anti-commute, 
$s^2 = \bar s^2 = \acom{s}{\bar s} = 0$.

As in the Abelian case, one defines a BRST charge $Q_{BRST}$ and anti-BRST charge $\bar Q_{BRST}$
analogous to \eqref{BRSTAbelian}. The space $\P$ of physical operators then is the
sector of vanishing ghost number of the BRST cohomology as in \eqref{physOP}. Unphysical states of indefinite norm are associated with BRST quartets.  To all orders in perturbation theory these do not contribute to the physical
scattering matrix and cannot be created from physical states by physical operators \cite{Becchi:1975nq, Kugo:1979gm}.  The
longitudinal gauge field, ghost $c$, anti-ghost $\bar c$ and NL field $b$ again form the elementary BRST
quartet\cite{Kugo:1979gm,Peskin:1995ev}. Contrary to the Abelian case, the transverse gluon of a non-Abelian gauge theory is  
part a non-perturbative BRST quartet \cite{Nakanishi:1990qm,Alkofer:2011pe} and not physical.

In the canonical formulation, the global color symmetry of this theory implies the conserved Noether currents,
\begin{align} 
j_\mu^{LCG\,a} &= \cp{A_\nu}{(F_{\nu\mu} + i \delta_{\mu\nu} b)}{a} \nonumber\\
&\qquad - i\cp{c}{\partial_\mu\bar c}{a} + i\cp{\bar c}{D_\mu c}{a} \, . \label{lcg_current}
\end{align}
Up to the gluonic QEoM,
\be 
J_\mu^a = - \partial_\nu F_{\nu\mu}^a + i s \bar s A_\mu^a
       = - \partial_\nu F_{\nu\mu}^a+i s (D_\mu \bar c)^a \, ,\label{ko_curr}
\ee
is equivalent to $j_\mu^{LCG\,a}$ and the color charges $\mathcal G^a$ and $\N^a$ in
\be\label{colorQ}
Q^a=\int d^3x j_0^{LCG\,a}\equiv \mathcal G^a+\N^a\ ,
\ee  
defined by,
\begin{subequations}
\begin{align}\label{defGN}
{\mathcal G}^a&:=-\int d^3x \partial_\nu F_{\nu 0}^a=\int_{S_\infty} d \sigma_i F_{0i} \, , \\
\N^a&:=i \acom{Q_{BRST}} {\int d^3x\, (D_0\bar c)^a} \nonumber\\
    &= \acom{\bar Q_{BRST}}{\int d^3x\, \pi_{\bar c}}\ ,
\end{align}
\end{subequations}
are individually conserved. Here $\pi_{\bar c}$ is canonically conjugate to the anti-ghost $\bar c$. Along the lines of  the argument in Abelian gauge theories following \eqref{Qfalse}, Kugo and Ojima showed 
 \cite{Kugo:1979gm, Kugo:1995km} that all physical operators are colorless and commute with $Q^a$ if two conditions are fulfilled: 
\begin{itemize}
\item[i)]  The Lorentz invariant function $f_{LCG}$ defined by the 
 transverse correlation function,   
\be\label{masslessVB}
\vev{ A^a_\sigma(y)\,  F^b_{\nu\mu}(x)}_\F= - i \delta^{ab}(\delta_{\sigma\mu} p_\nu-
\delta_{\sigma\nu}p_\mu) \frac{f_{LCG}(p^2)}{p^2}\ ,
\ee
must vanish at $p^2=0$, implying the absence of a massless vector boson in the adjoint 
representation of the group. 
\item[ii)] The function $u_{LCG}$, defined by,
\begin{multline}
-i\vev{A_{\sigma}^a(y)\,s\bar s A_\mu ^b(x) }= i\vev{(D_\sigma c)^a(y)\,(D_\mu \bar c )^b(x) }_\F 
 \\
= \delta^{ab}\left( T_{\sigma\mu}\,u_{LCG}(p^2)  -  L_{\sigma\mu}\right)\, , \label{ko_defu}
\end{multline}
must assume the value, $u(p^2\rightarrow 0)=-1$, in the infrared limit.
\end{itemize}

As in the Abelian case, condition (\ref{masslessVB}) with $f_{LCG}(0) = 0$, also holds  in a non-Abelian Higgs phase,
in which the vector bosons is massive. The infrared behavior of  $u_{LCG}(p^2)$ 
thus distinguishes between the Higgs and confinement phase in LCG. In terms  
QEoM of the gauge boson propagator this distinction may be reformulated as:
if  the Kugo-Ojima criterion is fulfilled, that is  if
\be\label{KOcrit}
u_{LCG}(0)=-1\ \ \text{and   }f_{LCG}(0)=0\ ,
\ee
the QEoM of the vector boson propagator, 
\begin{align}
  \delta^{ab}\delta_{\mu\sigma}\delta(x-y) &= \vev{A_{\sigma}^a(y)\,\var[S_{LCG}]{A_{\mu}^b(x)} } \\
  &  = \vev{A_{\sigma}^a(y)\,(-\partial_\nu F_{\nu\mu}^b - j_\mu^{LCG\,b}(x) } \nonumber
    \\& + i\vev{A_{\sigma}^a(y)\,s\bar s A_\mu ^b(x) }\,, \label{ko_dse} 
\end{align}
at vanishing momentum is saturated by \emph{unphysical} states only. 
By contrast, for $f_{LCG}(0)=0$ and 
$u_{LCG}(0)\neq -1$ the theory may describe a Higgs phase in which \emph{physical} states  contribute to the saturation of the transverse part of the current matrix 
element $ \vev{A_{\sigma}^a(y)\,  j_\mu^{LCG\,b})(x) }$ at vanishing momentum. Since one cannot be sure that physical states contribute when $u_{LCG}(0)\neq -1$, the criterion of Kugo and Ojima of \eqref{KOcrit} is a \emph{sufficient} criterion for confinement\cite{Kugo:1995km}. 

The longitudinal part of the correlation function in \eqref{ko_defu} is uniquely
determined by the equation of motion of the ghost field. It saturates the longitudinal part of
\eqref{ko_dse} for all momenta and implies that the current matrix element in LCG is transverse. 

If the Kugo-Ojima criterion is fulfilled,  the matrix element of the BRST-exact part of the (generalized) color current saturates the transverse gluonic QEoM of LCG at low momentum.  Assuming the BRST-symmetry is unbroken,  Kugo and Ojima proved that the physical sector in this case is colorless \cite{Kugo:1979gm, Kugo:1995km,Nakanishi:1990qm}. In a Higgs phase, the physical, albeit massive, vector boson would contribute to the 
transverse part of the gluonic QEoM at vanishing momentum and it would not be saturated by unphysical states only.  

Since color confinement is not compatible with physical states in the adjoint color representation, an alternative criterion for the confining phase, but one that includes the essential information,  is the saturation of the gluonic QEoM by
unphysical degrees of freedom in the infrared. We now investigate whether this proposal can be applied to a wider class of gauges.     

\subsection{Saturation and Confinement in Generalized Linear Covariant Gauges (GLCG) \label{sec_cc}}
Let us therefore next examine saturation of the gluonic QEoM in Generalized Linear 
Covariant  gauges (GLCG) given by the Lagrangian \cite{ThierryMieg:1985yv,Alkofer:2003jr} 
\begin{align}
\mathcal L_{GLCG}  &=  \LYM +\sa\left(\bar c^a \left(\frac{\xi}{2} b^a- i\partial_\mu A^a_\mu
\right)\right)\label{L_cc}\\
&=\LYM + \frac \xi 2 b^2 - i \partial_\mu A_\mu^a b^a - i  \alpha \left(D_\mu \bar c\right)^a
\partial_\mu c^a  \nonumber \\
    & - i \bar\alpha \partial_\mu \bar c^a \left(D_\mu c\right)^a + 
    \frac{\alpha \bar \alpha \xi}{2} \cp{\bar c}{c}{2}  \, , \nonumber
\end{align}
with $\alpha + \bar  \alpha = 1$. 
The Lagrangian of \eqref{L_cc} interpolates between LCG  
($\alpha = 0$), its Faddeev-Popov conjugate ($\alpha = 1$), and the ghost-antighost symmetric gauge 
at $\alpha=$ $\bar\alpha = \frac{1}{2}$. The generalized Kugo-Ojima confinement scenario for this Lagrangian is discussed in \cite{maderinprep}. For any value of the gauge parameters $\alpha$ and $\xi$, 
$\mathcal L_{GLCG}$  is globally color symmetric and invariant under the nilpotent BRST and anti-BRST transformations,
\begin{align}\label{brs_gag}
      &\sa A_\mu^a  = D_\mu c^a \, , \nonumber\\  
      &\sa c^a  = -\frac{1}{2} \cp{c}{c}{a} \, , \\
      &\sa \bar c  = b^a - \alpha \cp{\bar c}{c}{a}\, , \nonumber\\  
      &\sa b^a = -\alpha \cp{c}{b}{a} + 
      \frac{\alpha \bar\alpha}{2} \cp{\bar c}{\cp{c}{c}{}}{a} \, ;\nonumber
\end{align}      
\begin{align}
      &\sab A_\mu^a  = D_\mu\bar c ^a\, , \nonumber\\  
      & \sab \bar c^a  = -\frac{1}{2} \cp{\bar c}{\bar c}{a} \, , \nonumber\\  
	\label{abrs_gag}
      &\sab c  = - b^a - \bar\alpha \cp{\bar c}{c}{a}\, , \\  
      &\sab b^a =        - \bar \alpha \cp{\bar c}{b}{a} + 
      \frac{\alpha\bar\alpha}{2}\cp{\cp{\bar c}{\bar c}{}}{c}{a} 
       \,.\nonumber
\end{align}
The ghost-antighost symmetric gauge $\alpha=\bar\alpha=\frac{1}{2}$ possesses an additional 
continuous global $SL(2,R)$ symmetry generated by $\Pi_{\mp}$ and the ghost number $\Pi$,
\begin{multline}\label{SL2}
\Pi_+=\int d^4x \; c(x)\frac{\delta}{\delta \bar c(x)}\,,\\ \Pi_-=\int d^4x\; 
\bar c(x)\frac{\delta}{\delta c(x)}\,,\ \ \  [\Pi_+,\Pi_-]=2 \Pi\ .
\end{multline}
One verifies that the BRST and anti-BRST charges $Q_{BRST}$ and 
$\bar Q_{BRST}$, which generate the transformations of \eqref{brs_gag} and \eqref{abrs_gag}, anti-commute \cite{ThierryMieg:1985yv}, and that the graded 
algebra of $\{Q_{BRST},\bar Q_{BRST}, \Pi_+,\Pi_-,\Pi\}$ closes in ghost-antighost symmetric 
gauges. The space of physical operators $\P$ is defined as in \eqref{physOP}.

Due to the invariance of \eqref{L_cc} under global color transformations, the 
Noether currents,  
\begin{multline}
    j_\mu^{GLCG\,a} = \cp{A_\nu}{(F_{\nu\mu} + i b \delta_{\mu\nu})}{a} 
      - i \cp{c}{(\alpha D_\mu\bar c + \bar \alpha \partial_\mu \bar c)}{a} \\ + 
      i \cp{\bar c}{(\bar \alpha D_\mu c + \alpha \partial_\mu c )}{a}\, , \label{cc_cur}
\end{multline}
are conserved. These currents depend explicitly on the gauge parameter $\alpha$. They are part of the gluonic QEoM, which here takes the form,
\begin{align}
 \delta^{ab}_{\mu\sigma}\delta(x-y) & = \vev{A_{\sigma}^a(y)\,\var[S_{GLCG}]{A_{\mu}^b(x)} } 
 \label{cc_dse} \\
    & =- \vev{A_{\sigma}^a(y)\,(\partial_\nu F_{\nu\mu} + j_{\mu}^{GLCG})^b(x) } \nonumber\\
    & + \vev{A_{\sigma}^a(y)\,(i\sa\sab A_\mu )^b(x) }\,. \nonumber
\end{align}

The last term in \eqref{cc_dse} again involves only unphysical excitations and is of the same form as in the LCG studied above,
\begin{multline} 
\vev{A_{\sigma}^a(y)\,(i\sa\sab A_\mu )^b(x) }_\F \\= -i\vev{(D_{\sigma}c)^a(y)\,
(D_\mu \bar c)^b(x)}_\F \\ = \delta^{ab}\left( L_{\sigma\mu}-T_{\sigma\mu}\,u_{GLCG}(p^2) 
\right)\,.\label{cc_sat}
\end{multline}
The equation of motion of the ghost by itself does not suffice to determine 
the longitudinal part of \eqref{cc_sat}. Instead one has
\be\label{longcc}
i(\partial_\mu D_\mu \bar c)^a=\frac{\delta S_{GLCG}}{\delta c^a}- 
\frac{\xi \bar \alpha}{2} s (\bar c\times\bar c)^a\ .
\ee
Since $\vev{A_\mu^a} = 0$ and the BRST transformation is nilpotent,  \eqref{longcc} determines the 
longitudinal part of the correlation function in \eqref{cc_sat}. As in LCG, unphysical degrees of freedom saturate the longitudinal part of the gluonic QEoM in \eqref{cc_dse}, and the current matrix  element is transverse.

 The form factor, $f_{GLCG}(p^2)$, is defined as in \eqref{masslessVB}, and the same
discussion as in Sec.~\ref{sec_ko}  applies. The theory describes a Coulomb phase with a massless vector boson only if $f_{GLCG}(0)\neq0 $. The transverse part of the correlator in \eqref{cc_sat} defines a form factor $u_{GLCG}(p^2)$ whose value in the infrared can be used as a criterion in these gauges.  The transverse gluonic QEoM is saturated by
unphysical degrees  of freedom in the infrared and the phase is confining if $u_{GLCG}(0)=-1$. One thus formally has the same confinement criterion as in
LCG \cite{maderinprep}. However, although the unphysical correlation functions in \eqref{cc_sat} and \eqref{ko_defu} formally look similar, unphysical correlations differ, and in general $u_{GLCG}(p^2)\neq u_{LCG}(p^2)$ if $\alpha\neq 0$. The confinement
criterion stated above asserts that these functions in the confining phase may coincide at $p^2=0$ in any gauge parametrized by
$(\alpha,\xi)$. We next turn to covariant gauges that explicitly break
global color invariance to the Cartan subgroup.     

\subsection{Saturation and Confinement in  Maximal Abelian Gauges (MAG)\label{sec_mag}}
An equivariant BRST construction allows one to partially localize a non-Abelian gauge theory to an
equivalent Abelian model with the same gauge invariant correlators. This partial localization is
possible on the lattice \cite{Schaden:1998hz,Golterman:2004qv,Golterman:2012ig} as well as in the
continuum \cite{Schaden:1999ew,Golterman:2005ha,Ferrari:2013aza} and defines the continuum theory as the critical
limit of a lattice model with the same global symmetries.  The equivariant construction may be viewed
as a partial gauge fixing that leaves the Abelian Cartan subgroup free, hence Maximal Abelian Gauge. 
The resulting  Abelian gauge theory has the same gauge invariant physical correlation
functions as the non-Abelian model and the residual Abelian gauge symmetry of MAG can be dealt with as in any Abelian gauge theory. An $SU(N)$ gauge theory in MAG presents itself as an Abelian $(\uone)^{N-1}$ gauge
theory that is asymptotically free but retains typical Abelian Ward identities. In the following we
investigate how a confining phase  may manifest itself in this Abelian gauge theory. 

In MAG one discriminates between the Cartan subgroup and  the coset space. For an $SU(N)$ gauge
theory the mutually commuting (Hermitian) generators of the Cartan subgroup will be denoted by 
$\{T^i; [T^i,T^j]=0\ \forall i,j=1,\dots,N-1\}$, whereas the remaining $N(N-1)$ coset generators
carry Latin indices from the beginning of the alphabet $\{T^a; a=1,\dots,N(N-1)\}$. The non-Abelian
$su(N)$ connection is decomposed as  $\mathcal A_\mu =A_\mu^i T^i+ B_\mu^a T^a$ and the field
strength tensor may similarly be decomposed into Cartan and coset components, $\mathcal F_{\mu\nu} =
f_{\mu\nu}^i T^i + F_{\mu\nu}^{a}T^a$ with
\begin{subequations}
\label{Fdecomp}
    \begin{align}
     f_{\mu\nu}^i & = \partial_\mu A_\nu^i - \partial_\nu A_\mu^i + \cp{B_\mu}{B_\nu}{i} 
     \quad\text{ and } \label{mag_fsf}\\
     F_{\mu\nu}^a & = (\D_\mu B_\nu)^a - (\D_\nu B_\mu)^a + \cp{B_\mu}{B_\nu}{a}\,. \label{mag_fsF}
    \end{align}
\end{subequations}
Here and in the following the covariant derivative with respect to the Cartan gluons in the 
adjoint representation is denoted by $\D_\mu^{ab}$, see App.~\ref{app_conv}. 
The $SU(N)$ Yang-Mills Lagrangian in the components of  \eqref{Fdecomp} reads
\be 
    \LYM = \frac{1}{4} f_{\mu\nu}^i f_{\mu\nu}^i +  \frac{1}{4} F_{\mu\nu}^a F_{\mu\nu}^a\,.
    \label{mag_ym}
\ee

Although a more general discussion is possible, we for simplicity consider the gauge group $SU(2)$ in
the following. It illustrates our main points and connects to our
considerations in Sec.~\ref{sec_qed}. The Cartan subalgebra in this case is one dimensional and we suppress its index. The coset space is two dimensional with components $a=1,2$. 

The distinction between Abelian and coset degrees of freedom is accomplished by the ``gauge fixing"
part of the MAG Lagrangian,
      \begin{align}
	  \LGF^{MAG}&= \frac{i}{2} \se\seb \left( B_\mu^a B_\mu^a - i \lambda\bar c ^a c^a \right)
	  \label{mag_gf}\\
	  &  =  \frac{\lambda}{2} \eta^a \eta^a - i \eta^a(\D_\mu B_\mu)^a - i (\D_\mu\bar c)^a 
	  (\D_\mu c)^a  \nonumber \\
	  &  + i \cp{B_\mu}{\bar c}{} \cp{B_\mu}{c}{} + \frac{\lambda }{2}\,\cp{\bar c}{c}{2} \, .
	  \nonumber
      \end{align}
      $\LGF^{MAG}$  is obtained using the  equivariant BRST and anti-BRST transformations
\cite{Schaden:1998hz,Schaden:1999ew},
\begin{subequations} 
\label{eBRST}
     \begin{align}
	\se A_\mu & = \cp{B_\mu}{c}{}, & \seb A_\mu & =\cp{B_\mu}{\bar c}{}, \label{mag_eBRST_A} \\
	\se B^a_\mu & = (\D_\mu c)^a, & \seb B^a_\mu & = (\D_\mu \bar c)^a,  \label{mag_eBRST_B}\\
	\se c^a & = 0, & \seb \bar c^a & = 0, \label{mag_eBRST_c}\\
	\se \bar c^a &= \eta^a, & \seb c^a &= -\eta^a, \label{mag_eBRST_cb}\\
	\se \eta^a &= \frac{1}{2} \cp{\bar c}{\cp{c}{c}{}}{a}, & \seb \eta^a &= \frac{1}{2} 
	\cp{\cp{\bar c}{\bar c}{}}{c}{a} \label{mag_eBRST_eta},
      \end{align}
\end{subequations}
which generate infinitesimal gauge transformations in the coset space $SU(2)/U(1)$ with 
parameters $c^a(x)$ and $\bar c^a(x)$. $\LYM$ and covariantly coupled matter 
are invariant under these transformations.  Contrary to LCG, $SU(2)$ in MAG has two (anti-)ghosts only. 
The Lagrangian
\be\label{equivA}
\L^\uone_{MAG} = \LYM + \LGF^{MAG}
\ee
defines an $\uone$  invariant gauge theory. It includes exotic \uone-charged 
$B^a_\mu, c^a$ and $\bar c^a$ ``matter" and is invariant under the equivariant BRST and 
anti-BRST transformations of \eqref{eBRST} and under infinitesimal local $\uone$ transformations,
\begin{multline}
A_\mu\rightarrow\partial_\mu\theta,\quad B_\mu^a\rightarrow \cp{B_\mu}{\theta}{a},\quad c^a
\rightarrow \cp{c}{\theta}{a},\quad \\
    \bar c^a \rightarrow \cp{\bar c}{\theta}{a}\,\text{and}\quad \eta^a\rightarrow \cp{\eta}
    {\theta}{a}\,.\label{mag_uonetrafo} 
\end{multline}

Analogous to \eqref{BRSTAbelian} one can define the equivariant BRST and anti-BRST charges, $Q_{\varepsilon BRST}$ and $\bar Q_{\varepsilon BRST}$. The Lagrangian (\ref{mag_gf}) and the algebra of $Q_{\varepsilon BRST}$ and $\bar Q_{\varepsilon BRST}$ are ghost-antighost symmetric. In addition to the ghost number $\Pi$, one thus has the complete $sl(2,R)$ algebra of charges  typical of ghost-antighost symmetric gauges. However, in MAG the $\Pi_\mp$ of \eqref{SL2} involve only two coset (anti-)ghosts rather than the three of an  $SU(2)$ gauge theory in ghost-antighost symmetric GLCG.  

The equivariant BRST transformations (\ref{eBRST}) are \emph{not} nilpotent. $\se^2$, $\seb^2$ and
$\frac{1}{2}\acom{\se}{\seb}$ generate \uone \ transformations with a bosonic parameter $\theta(x) = \tfrac{1}{2}
\cp{c}{c}{}, \tfrac{1}{2} \cp{\bar c}{\bar c}{}\,\text{and}\,\frac{1}{2}\cp{\bar c}{c}{}$,
respectively. However,  $\se^2\Op = \seb^2\Op = \acom{\se}{\seb}\Op = 0$ for any operator $\Op$ 
invariant under the \uone \ transformations of \eqref{mag_uonetrafo}. The equivariant BRST algebra thus 
reduces to the usual  nilpotent BRST algebra on the set of \uone \ invariant functionals only. 

The residual \uone \ symmetry of the Lagrangian defined by \eqref{equivA} can be fixed to any gauge. To allow for comparison with Sec.~\ref{sec_qed} we here choose a linear covariant gauge
 \be\label{mag_uone}
      \LGF^\uone =  \frac{\xi}{2}b^2-ib\partial_\mu A_\mu 
 \ee
 with gauge-fixing parameter $\xi$ and (uncharged)  auxiliary field $b$.  
 $b$ is taken to be invariant under $\se$ and $\seb$, $\se b = \seb b = 0$. 

We thus consider the $SU(2)-$ Yang-Mills theory in MAG specified by the action 
$\Smag = \int\!d^4x \L_{MAG}(x)$ with,
      \be  \L_{MAG} = \LYM + \LGF^{MAG} + \LGF^\uone\, \label{mag_Lag}\,.\ee    
     
The $\uone$ gauge fixing $\LGF^\uone$ of \eqref{mag_Lag} not only explicitly breaks the local
\uone-gauge symmetry but global symmetries as well. $\LGF^\uone$ is
symmetric under global \uone-transformations, but breaks the global equivariant BRST, anti-BRST
and $SL(2)$ symmetries explicitly. For any, not necessarily local, operator $\Op$ one has the
Ward identities,
      \begin{subequations}
      \label{mag_wardId}
	  \begin{alignat}{2}
	    \label{U1Ward}
	      \vev{\delta_x \Op}&=\vev{\Op\ \delta_x \Smag}& &=i\vev{\Op \ \partial^2 b(x)}\\
	    \label{seWard}
	      \vev{\se\Op}&=\vev{\Op\ \se \Smag}& &=i\vev{\Op\  \int (\partial_\mu b)\se A_\mu}\\
	    \label{sebWard}
	      \vev{\seb\Op}&=\vev{\Op\ \seb \Smag}& &=i\vev{\Op\  \int (\partial_\mu b)\seb A_\mu}\ , 
	  \end{alignat}
      \end{subequations}
    where,
\begin{multline}
	 \label{delta}
	  \delta_x \Op=\Biggl(\partial_\mu\var{A_\mu(x)} +  B_\mu(x)\times\var{B_\mu(x)} + 
	  c(x)\times\var{c(x)} \\
	    + \bar c(x)\times\var{\bar c(x)} + \eta(x)\times\var{\eta(x)}\Biggr)\Op,
\end{multline}
generates a local $\uone$ gauge transformation at $x$.  Defining the set  $\W$ of   
\uone-invariant operators,
\be
	\label{mag_U1inv}
	  \Op\in \W\Leftrightarrow \delta_x \Op=0\ ,
	\ee    
we show in App.~\ref{app_U1} that
\be\label{invOp}
\vev{\se\Op}=\vev{\seb\Op}=0\ \ \text{for all  } \Op\in\W\ .
\ee 
On the set $\W$ of \uone \ invariant operators  of the equivalent \emph{Abelian} 
gauge theory one thus recovers $\se$ and $\seb$ as nilpotent BRST symmetries. 
One then can define the set of physical operators $\P\subseteq\W$ of 
the underlying non-Abelian $SU(2)$ gauge theory by the equivariant cohomology,
\begin{multline}\label{physOp_equi}
\P= \{ \O\in\W; \com{ Q_{\varepsilon BRST} }{\O}=0\ \text{and } [\O,\Pi]=0 \} \\ 
/ \{ \com{Q_{\varepsilon BRST} }{\O} ;\O\in\W \ \text{and}\  [\O,\Pi]=\O \}\ .
\end{multline} 
The conserved \uone \ current of the Cartan subalgebra of $SU(2)$ in MAG is,
\begin{subequations}
    \begin{align}
	j^{MAG}_\mu &=  \cp{B_\nu}{F_{\nu\mu}}{} + i \cp{B_\mu}{\eta}{} + 
	i \cp{\bar c}{D_\mu c}{}\nonumber \\ 
		    & \hspace{30mm} - i \cp{D_\mu \bar c}{c}{} \,, \label{mag_curr}
	\intertext{and can be rewritten in the form,}
	j^{MAG}_\mu & = i \com{\se}{\seb}A_\mu +B_\nu\times( F_{\nu\mu} - i \delta_{\mu\nu} \eta) 
	\in \W\label{mag_curr_brst}\ .
    \end{align}
\end{subequations}
 In fact, each term in \eqref{mag_curr_brst} separately is an element of $\W$, 
 but $ i \com{\se}{\seb}A_\mu$ does not create \emph{physical} states.

The QEoM of the Cartan gluon propagator depends on the conserved Abelian Noether current of 
\eqref{mag_curr_brst} as in the Abelian gauge theory studied in Sec.~\ref{sec_qed},
\begin{align} 
\delta_{\sigma\mu}\delta(x-y) & = \vev{A_{\sigma}(y)\, \var[\Smag]{A_{\mu}(x)}} \label{mag_dse}\\
  &= -\vev{A_\sigma(y)\,\left(  \partial_\nu f_{\nu\mu} + j^{MAG}_\mu \right)(x)}  \nonumber\\
  & + \vev{A_\sigma(y)\,i\partial_\mu b(x)} \,.\nonumber
\end{align}
As for an unbroken Abelian gauge theory,  the last term of \eqref{mag_dse} saturates the 
longitudinal part of the gluonic  QEoM because of the Abelian Ward identity of \eqref{U1Ward},
\begin{multline}\label{satlongMAG} 
\partial_\sigma \delta(x-y)=\vev{\delta_x A_\sigma(y)}=i\vev{A_\sigma(y)\, \partial^2 b(x)} \\ 
	\Rightarrow \vev{A_\sigma(y)\, i\partial_\mu b(x)}_\F=L_{\sigma\mu}\ .
\end{multline}

The first correlator in \eqref{mag_dse} is transverse due to the anti-symmetry of 
$f_{\mu\nu}$ and the current matrix element thus is transverse as well,
\be\label{jmagsat}
 \vev{A_\sigma(y)\,j^{MAG}_\mu(x)}_\F=(f_{MAG}(p^2)-1)T_{\mu\nu}\ .
\ee

The functions $f_{MAG},\, u_{MAG},\, h_{MAG},$ and $\ell_{MAG}$ are defined from the 
correlators
\begin{subequations}
\be
 -\vev{A_\sigma(y)\, \partial_\nu f_{\nu\mu} (x)}_\F  = f_{MAG}(p^2) T_{\sigma\mu} 
 \label{mag_deff}
\ee
 \begin{multline}
 i\vev{A_\sigma(y)\, \com{\se}{\seb}A_\mu (x)}_\F  = u_{MAG}(p^2) T_{\sigma\mu} \\ 
 + \ell_{MAG}(p^2) L_{\sigma\mu} \label{mag_defi}
\end{multline}
\begin{multline}
 -\vev{A_\sigma(y)\, \cp{B_\nu}{(F_{\nu\mu} - \delta_{\mu\nu} i \eta)}{}}_\F  
 = h_{MAG}(p^2) T_{\sigma\mu} \\ +\ell_{MAG}(p^2) L_{\sigma\mu} \,,
\end{multline}
\end{subequations}
and the transverse part of \eqref{mag_dse} yields the constraint, 
\be\label{MAGcond}
f_{MAG}(p^2) + h_{MAG}(p^2) = 1+ u_{MAG}(p^2)\ .
\ee
As in LCG and GLCG, the transverse gluonic QEoM is saturated by unphysical degrees of freedom and 
the Cartan color charge of physical states vanishes when the
the Kugo-Ojima-like criterion
\be\label{MAG_sat}
f_{MAG}(0)=0 \ \text{and}\ \ u_{MAG}(0)=-1\, ,
\ee
holds since it implies that $h_{MAG}(0)=0$. The conditions~(\ref{MAG_sat})  guarantee saturation of the gluonic  QEoM
in the infrared by unphysical degrees of freedom in MAG and implies that physical states   are colorless. \eqref{jmagsat} together with
\eqref{MAG_sat} imply that this scenario can only be realized in MAG if unphysical degrees of freedom created by the  \emph{conserved Abelian current} $j_\mu^{MAG}$ saturate the QEoM of the Abelian propagator 
at low momenta.  From the point of view of the Abelian gauge theory, saturation of the transverse equation at  low momenta  in confinement and Higgs phases thus is similar. In MAG the only difference is that whereas some \emph{physical} degrees of freedom contribute to the matrix element of the current at vanishing momentum  in the Higgs phase, only \emph{unphysical} states contribute in the confinement phase.
The current saturates the QEoM of the Abelian propagator at low energies in the Higgs phase described by \eqref{LH} as well as in the confinement phase of the $SU(2)$  gauge theory in MAG  defined by \eqref{equivA}. This supports the idea that the phases and the two Abelian models describing them are dual\cite{'tHooft:1979bi}. 

In this context it is interesting to consider the condition $f_{MAG}(0) = 0$ more closely. 
The analogous condition implies a massive \emph{physical} vector boson in the Abelian theory 
described by \eqref{LHiggs}. The Cartan gluon on the other hand is not a physical asymptotic 
state in the confinement phase and it has been conjectured \cite{Huber:2009wh} that the Abelian 
propagator in this case may even be enhanced in the infrared.  That this scenario can be 
reconciled with the criterion of \eqref{MAG_sat} rests on the definition 
\eqref{mag_fsf} of the Abelian field strength tensor $f_{\mu\nu}$.  \eqref{mag_deff} implies  
\begin{multline} 
f_{MAG}(p^2) T_{\sigma\mu} = p^2 T_{\mu\nu} \vev{A_\sigma(y) \,A_\nu(x)}_\F \\-  
\vev{A_\sigma(y) \ \partial_\nu\cp{B_\nu}{B_\mu}{}\!(x)}_\F \,. \label{mag_satf}
\end{multline}
Although a massive Abelian vector boson allows one to fulfill $f_{MAG}(0)=0$, 
the last correlator of \eqref{mag_satf} prohibits one from asserting that the 
diagonal gluon propagator \emph{has to be} suppressed at low momenta. 

Introducing the function $\alpha_{MAG}(p^2)$, 
\be\label{enhanced}
\vev{A_\sigma(y) \ \cp{B_\nu}{B_\mu}{}\!(x)}_\F=-i (\delta_{\sigma\nu} p_\mu
-\delta_{\sigma\mu}p_\nu) \alpha_{MAG}(p^2) ,
\ee
\eqref{mag_satf} states that,
 \be f_{MAG}(p^2)  = p^2 ({\textstyle \frac{1}{3}}T_{\mu\nu} \vev{A_\mu(y) \,A_\nu(x)}_\F - 
  \alpha_{MAG}(p^2)) \,. \label{mag_satf1}
  \ee
If the Cartan gluon  correlator is infrared enhanced, \eqref{mag_satf1} determines only the 
infrared singular behavior of  $\alpha_{MAG}(p^2)$  when $f_{MAG}(0)=0$.

To gain some more information about these functions, we define a \uone invariant transverse field strength,
\be\label{EM}
G_{\mu\nu}=\partial_\mu A_\nu-\partial_\nu A_\mu\,.
\ee
It is not an invariant of the equivariant BRST (or anti-BRST) and,  in contrast to the $\uone$ gauge 
theory considered in Sec.~\ref{sec_qed},
is not a physical operator of the $SU(2)$ gauge theory. The function $u_{MAG}(p^2)$ defined in 
\eqref{mag_defi} also describes the correlation functions,
\begin{multline}
\label{uG}
 u_{MAG}(p^2)(\delta_{\rho\mu}p_\sigma-\delta_{\sigma\mu}p_\rho) =  \vev{G_{\rho\sigma}(y)\, 
 \com{\se}{\seb}A_\mu (x)}_\F \\
 =\vev{\seb G_{\rho\sigma}(y)\, \se A_\mu (x)}_\F- \vev{\se G_{\rho\sigma}(y)\, 
 \seb A_\mu (x)}_\F \, ,
\end{multline}
where \eqref{invOp} was used since  $\{G_{\mu\nu},\se A_\mu,\seb A_\mu\}\subset\W$. 
Using the definition in \eqref{EM} and exploiting Poincar{\'e} invariance, \eqref{uG} implies,
\begin{multline}\label{MAG_cc}
u_{MAG}(p^2) T_{\mu\nu}+v_{MAG}(p^2) L_{\mu\nu}=2i\vev{\seb A_\nu(y)\,\se A_\mu(x)}_\F \\ 
=2i\vev{ \cp{B_\nu}{\bar c}{}(y)\, \cp{B_\mu}{c}{}(x)}_\F\ ,
\end{multline}
in close analogy to \eqref{ko_defu} in general covariant gauges. Neither \eqref{uG} nor the 
equations of motion constrain the longitudinal function $v_{MAG}(p^2)$ in this case. 
Since \eqref{invOp} holds only for \uone-invariant functionals in $\W$, $v_{MAG}(p^2)$ 
also need not be related to $\ell_{MAG}(p^2)$ defined by 
\eqref{mag_defi}\footnote{Note that the definition of the function $u_{MAG}(p^2)$ in 
MAG apparently differs  by a factor of $(-2)$ from that of GLCG given by \eqref{cc_sat}, 
since $(D_\mu c)^3= \partial_\mu c^3+\cp{B_\mu}{c}{3}$ formally differs from 
$ \cp{B_\mu}{c}{}$ in \eqref{MAG_cc} by a longitudinal contribution only. However, 
\eqref{cc_sat} and \eqref{MAG_cc} are gauge dependent correlation functions that are not 
required to coincide in two different gauges. In addition, as LCG and MAG are not analytically connected, \cite{Hata:1992np}, no quantitative relation should be expected between the functions $u_{LCG}$ and $u_{MAG}$.}.

We next study signatures of the confining phase in a gauge that is not covariant.

\subsection{Saturation and Confinement in  Non-Abelian Coulomb Gauge (CG)\label{sec_CG}} 
Coulomb gauge breaks manifest Lorentz covariance by treating timelike and spacelike gluons 
differently. We here study to what extent the Kugo-Ojima criterion depends on a manifestly 
Lorentz-invariant gauge condition. Coulomb gauge is described by the Lagrangian,
\begin{align}
 \L_C &= \LYM -i s\left(\bar c^a \partial_i A_i^a \right)
  \label{c_lag}\\
      &= \LYM - i b^a \partial_i A_i^a - i \partial_i\bar c^a (D_ic)^a \,,
       \nonumber
\end{align}
where Latin indices denote spatial components of a Lorentz-vector, $i,j,\dots = 1,2,3$. 
The BRST transformations are the same as in \eqref{ko_BRST}, 
 \begin{align}\label{C_BRSTtrafo}	
      s A_0^a  = (D_0 c)^a, && s A_i^a  = (D_i c)^a, && sc^a  = -\frac{1}{2} \cp{c}{c}{a},  
      \nonumber\\
      s \bar c ^a  = b^a, &&  sb^a = 0 \,.
 \end{align}
The BRST charge in Coulomb gauge can be written in terms of Gauss's law \cite{Zwanziger:1998ez}
    \be Q_{BRST} = - \int\! d^3x\, c^a (D_i F_{i0})^a = \int\! d^3x \,c^a \var[S_C]{A_0^a} 
     \,.\label{C_BRST}
     \ee
The anti-BRST transformations and corresponding charge may be defined analogously and the set of
physical operators again is given by the BRST cohomology of \eqref{physOP}. Coulomb gauge
 manifestly preserves global color symmetry and the color currents
\begin{subequations}
\begin{align}
\label{temporalj}
 {j^C_0}^a  &= \cp{A_i}{F_{i0}}{a},\\ 
{j^C_i}^a &= \cp{A_0}{F_{0i}}{a} + \cp{A_j}{F_{ji}}{a} + \cp{A_i}{b}{a} \nonumber \\ 
	    & - i \cp{c}{\partial_i \bar c}{a} + i \cp{\bar c}{D_i c}{a} \, ,
\label{spatialj}
\end{align}
\end{subequations}
are conserved. The absence of manifest Lorentz invariance in Coulomb gauge implies 
two distinct gluonic QEoMs. The QEoM of the time component reads
\begin{align} 
\delta^{ab} & = \vev{A_0^b(y)\,\var[S_C]{A_0^a(x)}}_\F \label{C_temporaleom}\\ 
    & = \vev{A_0^b(y)\, \left( -\partial_iF_{i0}^a(x)  - j^{C\,a}_0(x)\right) }_\F \nonumber\\
    &= -\vev{A_0^b(y)\,D_iF_{i0}^a(x) }_\F\,. \nonumber 
\end{align}
Since all physical states satisfy Gauss's Law in Coulomb gauge, this equation of motion is saturated
by unphysical states only, whether the model confines or not. To see that all states that  contribute to
\eqref{C_temporaleom} are unphysical note that physical states $\ket{\Psi_{phys}}$  are created by
physical operators defined in \eqref{physOP}. They have vanishing ghost number and are annihilated by
the ``Gauss-BRST" charge of \eqref{C_BRST},
\be Q_{BRST} \ket{\Psi_{phys}} = 0\,. \label{C_subs} 
\ee
The ghost field $c$ does not annihilate $\ket{\Psi_{phys}}$, since its only effect is to create a 
ghost. \eqref{C_subs} thus has to be ensured by gluonic contributions only, and one gets back Gauss's law as the subsidiary condition,
\be \var[S_C]{A_0^a(x)}\, \ket{\Psi_{phys}} = 0 \qquad \forall\,x\,. \label{C_subs2}
\ee
Any non-vanishing contribution to \eqref{C_temporaleom} thus must be due to unphysical  
$\ket{\psi} \not\in \{\ket{\Psi_{phys}}\}$. In App.~\ref{app_Coulomb} we give an 
explicit calculation of the rhs of \eqref{C_temporaleom}, and relate the propagator of the temporal 
gluon to the Faddeev-Popov operator to show that it is saturated by instantaneous contributions only.

The discussion of the spatial components of the gluonic QEoM is very similar to that in LCG. 
The equation of motion for the spatial part of the gluon propagator is given by
\begin{align}
 \delta^{ab}\delta_{ij} & = \vev{A_j^b(y)\,\var[S_C]{A_i^a(x)}}_\F \label{coul_spatialeom} \\
    &= -\vev{A_j^b(y)\, \left(  \partial_\nu F_{\nu i}^a(x) + j^a_i(x) \right) }_\F \nonumber \\
    & + i \vev{A_j^b(y)\,s(D_i\bar c) ^a}_\F\, .  \nonumber
\end{align}
The first matrix element necessarily is spatially transverse in Coulomb gauge with 
$\partial_i A_i=0$.  The Faddeev-Popov operator of Coulomb gauge is instantaneous,
\begin{align}
 M^{ab}(x,y) &=- \partial_iD_i^{ab}\delta(\textbf{y} - \textbf{x} ) \delta(y_0-x_0)
	\\ & := \delta(y_0-x_0) M^{ab}(\textbf x,\textbf y )\ , 
\end{align}
and the contribution of the last term in \eqref{coul_spatialeom} therefore is instantaneous,   
 \begin{multline}
 i \vev{A_j^b(y)\,s\bar s A_i^a} = -i \vev{(D_j c)^b(y)\,(D_i\bar c)^a(x)} \\ = 
 -i \vev{(D_j^{bc}(y) \,(D_i^{ad}(x)\left[M^{-1}(\textbf{y, \textbf x}) \right]^{cd}} 
 \delta(y_0-x_0)\label{coul_inst}\ . 
 \end{multline}
Its Fourier-transform depends on spatial momenta only. The QEoM of the 
ghost gives the longitudinal part of the correlation function
\be  -i \vev{(D_j c)^b(y)\,(D_i\bar c)^a(x)}_\F = - t_{ij} u_C(\textbf p ^2) + l_{ij}\,,
\ee
where $t_{ij}$ and $l_{ij}$ are the longitudinal and transverse spatial projectors. 
The confinement criterion in Coulomb gauge reads,
\be\label{Coulconf} \lim_{\textbf p ^2\rightarrow 0} u_C(\textbf p ^2) = -1 \qquad 
\text{ and} \qquad \lim_{\textbf p ^2\rightarrow 0} f_C(p_0,\textbf p ) = 0, 
\ee
where the function $f_C(p_0,\textbf p)$ is defined by
\be -\vev{A_k^b(y)\,\partial_\nu F_{\nu j}^a(x)}_\F = \delta^{ab}t_{kj} f_C(p_0,\textbf p)\, .
\ee
The conditions of \eqref{Coulconf} ensure that the spatial gluonic QEoM is saturated by unphysical degrees of freedom in Coulomb gauge.

It is interesting that the correlation function in \eqref{coul_inst} is related to the horizon 
function of minimal Coulomb gauge in a finite quantization volume $V$ 
\cite{Zwanziger:1988jt,Zwanziger:1989mf,Zwanziger:1992qr,Vandersickel:2012tz},
\begin{align}\label{horizonCoula}
 H(A)& \equiv -i \delta^{ab}\delta_{ij} \int\!d^3xd^3y \, \nonumber \\
&\qquad \times (D_j^{bc}(y) \,   (D_i^{ad}(x)\left[M^{cd}(\textbf{y, \textbf x}) \right]^{-1}  \,.
\end{align}
with        
\be
\label{horizonCoul}
\vev{H(A)}  = V (N_c^2 - 1) \lim_{\textbf{p}^2\rightarrow 0}(1-2 u_C(\textbf p ^2)) \,.
\ee
In minimal Coulomb gauge, the configuration space is constrained to the first Gribov region by imposing the constraint, 
 \be \label{horizon_cond}
 \vev{H(A)} = 3 V (N_c^2 - 1) \, , \ee 
which in fact is equivalent to the condition  $u_C(0) = -1$ for color confinement of 
\eqref{horizonCoul}. A similar relation between the Kugo-Ojima criterion and the horizon condition also holds in minimal Landau gauge \cite{Zwanziger:1992qr, Dudal:2009xh}, to which we now turn.

\subsection{Saturation and Confinement in the covariant Gribov-Zwanziger(GZ) theory \label{sec_GZ}}  To avoid summation
over gauge equivalent configurations, the GZ approach seeks to dynamically  restrict the path
integral to the first Gribov region  \cite{Gribov:1977wm,Zwanziger:1988jt,
Zwanziger:1989mf,Zwanziger:1992qr,Vandersickel:2012tz}.  The restriction leads to a horizon
condition similar to \eqref{horizon_cond} and  can be implemented in a local renormalizable field
theory with additional auxiliary ghosts.   It was shown \cite{Maggiore:1993wq,Vandersickel:2012tz}
that the   GZ Lagrangian differs from $\LYM$ by BRST exact terms only. An infrared analysis of the
GZ action reveals that its scaling solution coincides exactly with the solution calculated from the Faddeev-Popov action for Landau gauge  \cite{Huber:2009tx}: the ghost propagator is infrared enhanced and the gluon propagator
infrared suppressed, the respective infrared exponents are identical. 
This corroborates the argument that for functional equations it suffices to take into account the appropriate boundary conditions, and no explicit restriction in the path integral measure is
required. However, the horizon condition 
implies that the BRST symmetry of the GZ action is spontaneously broken. 
In this last section we want to investigate how the gluonic QEoM in minimal Landau gauge is saturated in the infrared, even though the spontaneously broken BRST symmetry prohibits a definition of physical operators as in the foregoing sections.

The auxiliary ghost-fields, 
$\phi^{a}_{\mu b},\bar\phi^{a}_{\mu b},\omega^{a}_{\mu b}, \text{and } \bar \omega^{a}_{\mu b}$ 
are vector fields with two color indices that transform under the adjoint representation of the 
global color group  (in $SU(3)$ they are reducible  
$\mathbf{8}\otimes\mathbf{8}=\mathbf{1}\oplus\mathbf{8}\oplus\mathbf{27} 
\oplus\mathbf{8}\oplus\mathbf{10}\oplus\overline{\mathbf{10}}$ color tensors),  
\be \delta\Psi^{a}_{\mu b} = g f^{acd} \Psi_{\mu b}^{c} \delta\vartheta^d +  
g f^{bcd}\Psi_{\mu c}^{a}\delta\vartheta^d \label{gz_traf2} \,
\ee
for any 
$\Psi^{a}_{\mu b} \in \{\phi^{a}_{\mu b},\bar\phi^{a}_{\mu b},\omega^{a}_{\mu b},
\bar \omega^{a}_{\mu b}\}$. While $\phi^{a}_{\mu b}$ and $ \bar \phi^{a}_{\mu b}$ 
are bosonic, $\omega^{a}_{\mu b}$ and $\bar \omega^{a}_{\mu b}$ are fermionic. 
The auxiliary ghosts form a BRST quartet,
\begin{align}
	s\phi^{a}_{\mu b} &= \omega^{a}_{\mu b}\, , & s\omega^{a}_{\mu b} & = 0, \\
	s\bar \omega^{a}_{\mu b} &= \bar\phi^{a}_{\mu b}\, , & s\bar \phi^{a}_{\mu b} & = 0 \,.
\end{align}
Including this auxiliary quartet in the BRST transformations of \eqref{ko_BRST}, 
the BRST exact part of the GZ Lagrangian is,
\be \L_{GZ}^{gf} = s(i\partial_\mu\bar c^a A^a_\mu + (\partial_\mu \bar \omega^{a}_{\nu b}) 
D_\mu^{ac}\phi_{\nu b}^{c} ) \label{gz_lag_before} \,.
\ee

The restriction of configuration space to the first Gribov region can be interpreted as a 
spontaneous
breakdown of this BRST symmetry. As in any instance of a spontaneously broken symmetry it 
is advantageous to express the Lagrangian in terms of fluctuations about the symmetry breaking 
ground state. In the GZ framework this amounts to a shift of the fields by
\begin{subequations}\label{gz_shift}
\begin{align}
	\phi^{a}_{\mu b}(x) & = \vph^{a}_{\mu b}(x) - \gamma^{1/2} x_\mu \delta^{a}_b \, ,\\
	\bar \phi^{a}_{\mu b}(x) & = \bar\vph^{a}_{\mu b}(x) + \gamma^{1/2} x_\mu 
	\delta^{a}_b \, , \\
	b^a(x) & = b^a(x) + i \gamma^{1/2}x_\mu \tr[a]{\bar \varphi_\mu}(x) \, ,\\
	\bar c^a(x) & = \bar c^a(x) + i \gamma^{1/2}x_\mu \tr[a]{\bar \omega_\mu}(x) \,,
\end{align}\end{subequations}
where $  \tr[a]{\Psi_\mu} = g f^{abc} \Psi^{b}_{\mu c}$.\footnote{Note that on a finite torus 
with \emph{antiperiodic} boundary conditions for the auxiliary ghosts, 
this $x$-dependent shift can be interpreted as quantization about a 
classical solution to the equations of motion.}
This change of variables in  
\eqref{gz_lag_before} gives the GZ Lagrangian of minimal Landau gauge 
\cite{Maggiore:1993wq,Vandersickel:2012tz},
\begin{align}
 \L_{GZ}^{gf}& = s(i\partial_\mu\bar c^a A^a_\mu + 
 (\partial_\mu \bar \omega^{a}_{\nu b}) D_\mu^{ac}\vph_{\nu b}^{c} - 
 \gamma^{1/2} D_\mu^{ac} \bar \omega^{c}_{\mu a} ) \nonumber \\
	& =  i \partial_\mu b^a A_\mu^a - i (\partial_\mu \bar c^a)(D_\mu c)^a + 
	(\partial_\mu \bar \vph^{a}_{\nu b})D_\mu^{ac} \vph_{\nu b}^{c} \nonumber \\
	& - (\partial_\mu \bar \omega^{a}_{\nu b})D_\mu^{ac}\omega_{\nu b}^{c}-  
	(\partial_\mu \bar \omega^{a}_{\nu b}) \cp{D_\mu c}{\vph_{\nu b}}{a} \label{gz_lag}\\
	&  + \gamma^{1/2} \left( D_\mu^{ac} (\vph^{c}_{\mu a} - \bar\vph^{c}_{\mu a}) - 
	\cp{D_\mu c}{\bar \omega_{\mu a}}{a}  \right) - \gamma d N_c  \,.\nonumber
\end{align}
Although the shift (\ref{gz_shift}) and the BRST transformations are $x$-dependent, 
the shifted Lagrangian (\ref{gz_lag}) does not include any explicit $x$-dependence, and 
is Poincar{\'e} invariant.

The  BRST variations of the shifted quantum fields are\footnote{This global transformation may appear to go outside the framework of  standard quantum field theory because of the large change at infinity.  However Noether's theorem and the Ward identities based on it rely for their validity on the infinitesimal local transformation $s_\epsilon = \epsilon(x) s$, that acts in particular on $\bar \omega ^{ab}_\mu$ according to
\be
\nonumber
s_\epsilon \bar \omega ^{ab}_\mu(x) = \epsilon(x)\bar \vph^{ab}_\mu(x) + \gamma^{1/2}\delta^{ab}
x_\mu \epsilon(x) .
\ee
Here $\epsilon(x)$ may be chosen to be zero outside a small but arbitrary region, so the transformation at large $x$ is strictly zero.  It is sufficient that the variation of the local Lagrangian under this infinitesimal and local transformation be proportional to $\partial_\mu \epsilon$, 
which it is, $s_\epsilon {\cal L}_{GZ} \sim j_\mu \partial_\mu \epsilon$.  The global transformation may be sidestepped \cite{Vandersickel:2012tz}.}
\begin{subequations}
\begin{align}
 sA_\mu^a & = (D_\mu c)^a \, , & sc^a&=-\frac 1 2 \cp{c}{c}{a}\label{sAc} \, ,\\ 
 s\bar c^a & = b^a\, ,& sb^a & = 0 \, , \label{sbcb}\\
 s\vph^{ab}_\mu & = \omega^{ab}_\mu \, , & s\omega^{a}_{\mu b} & = 0 \, , \label{sphiw}\\ 
 s \bar \omega ^{ab}_\mu(x) & = \bar \vph^{ab}_\mu(x) + \gamma^{1/2}\delta^{ab}x_\mu \, , & 
 s\bar\vph^{ab}_\mu & = 0 \label{sphibwb}\, .
\end{align}
\end{subequations}

The Gribov parameter $\gamma$ is found by demanding that the model is quantized about an 
extremum of the quantum effective action $\Gamma$,
\be \var[\Gamma]{\gamma} = 0\,. \ee
 The inhomogeneous term of \eqref{sphibwb} causes the BRST symmetry of the local Lagrangian to be
 spontaneously broken for any extremum of the quantum action with non-vanishing $\gamma$.  It is perhaps
 worth noting that for
 $\gamma\neq 0$ the Poincar\'e generators do not commute with the BRST charge even though Poincar\'e invariance is \emph{not} spontaneously broken.

To proceed with our investigation of confinement criteria in various gauges, we consider 
the gluonic QEoM  implied by the Lagrangian $\L_{GZ} = \LYM + \L_{GZ}^{gf}$
\begin{align}
    \var[S_{GZ}]{A_\mu^a} & = - \partial_\nu F_{\nu\mu}^a - j_\mu^{LCG\,a} + s(D_\mu \bar c)^a 
     \label{gz_eom1}  \\ 
      &+ \cp{\vph}{\partial_\mu \bar \vph}{a} + \cp{\omega}{\partial_\mu\bar\omega}{a} 
      - \cp{c}{\cp{\partial_\mu \omega}{\vph}{}}{a} \nonumber\\
      & - \gamma^{1/2} \cp{c}{\tr{\omega_\mu}}{a} + \gamma^{1/2} \tr[a]{\vph_\mu-\bar 
      \vph_\mu} \, , \nonumber
\end{align}
where we suppressed all indices that are summed. Contractions with structure constants in the 
``covariant" and ``contravariant" color indices are denoted by
\be
\cp{\Psi}{\Omega}{a} = g f^{acd} \Psi^{c}_{\mu b} \Omega^{d}_{\mu b} 
\ee
and
\be
\cpt{\Psi}{\Omega}{a} = g f^{acd} \Psi^{b}_{\mu c} \Omega^{b}_{\mu d} \,.
\ee
\eqref{gz_eom1} includes the global color current $j_\mu^{LCG\,a}$ of LCG given in 
\eqref{lcg_current}. However,  $j_\mu^{LCG\,a}$ is not the conserved color current of the 
GZ action since the auxiliary fields transform according to \eqref{gz_traf2}. 
The corresponding conserved color current is
\begin{align}
    j_\mu^{GZ\,a} & = j_\mu^{LCG\,a} + \cp{c}{\cp{\partial_\mu \bar \omega}{\vph}{}}{a} + 
    \gamma^{1/2}\cp{c}{\tr{\bar\omega_\mu}}{a}  \nonumber\\
	    &  - \cp{\vph}{\partial_\mu \bar \vph}{a} - \cpt{\vph}{\partial_\mu \bar \vph}{a} - 
	    \cp{\bar \vph}{D_\mu \vph}{a}   \nonumber \\
	    & - \cpt{\bar \vph}{D_\mu \vph}{a} - \cp{\omega}{\partial_\mu\bar \omega}{a} -
	     \cpt{\omega}{\partial_\mu\bar \omega}{a}\label{gz_cur} \\
	    &  + \cp{\bar \omega}{D_\mu \omega}{a} + \cpt{\bar \omega}{D_\mu \omega}{a}  
	    \nonumber \\
	    &  + \cp{\bar\omega}{\cp{D_\mu c}{\vph}{}}{a} + 
	    \cpt{\bar\omega}{\cp{D_\mu c}{\vph}{}}{a} \,. \nonumber
\end{align}
Using \eqref{gz_cur} the QEoM of \eqref{gz_eom1} may be  rewritten as,
\be \var[S_{GZ}]{A_\mu^a} = - \partial_\nu F_{\nu\mu}^a - j_\mu^{GZ\,a} + s\chi^a_\mu \,, 
\label{gz_eom2}
\ee
with 
\begin{multline}
 \chi^a_\mu = (D_\mu \bar c)^a - \cp{\bar \omega}{D_\mu \vph}{a}  - 
 \cpt{\bar \omega}{D_\mu \vph}{a}\\
  - \cpt{\vph}{\partial_\mu \bar \omega}{a} - \gamma^{1/2} \tr[a]{\bar\omega_\mu} \,. 
\end{multline}
The gluonic QEoM of the GZ action therefore has the now already familiar form
\begin{align}
   \delta^{ab}\delta_{\mu\sigma}\delta(x-y) &= \vev{A_{\sigma}^a(y)\,\var[S_{GZ}]{A_{\mu}^b(x)} }
    \label{gz_dse}    \\
   & = -\vev{A_{\sigma}^a(y)\,(\partial_\nu F_{\nu\mu}^b + j_\mu^{GZ\,b})(x) } \nonumber\\
   &  + i\vev{A_{\sigma}^a(y)\,s\chi^b_\mu(x) }\ . \nonumber
\end{align}

Color transport is short-ranged and the current matrix element does not contribute in the 
infrared if the functions $f_{GZ}(p^2)$ and $u_{GZ}(p^2)$ defined by
\begin{subequations}
\label{GZformfactors}
\begin{align}
-\vev{A_{\sigma}^a(y)\,\partial_\nu F_{\nu\mu}^b (x) }_\F&=T_{\sigma\mu} f_{GZ}(p^2) \label{fGZ}\\
i\vev{A_{\sigma}^a(y)\,s(\chi^b_\mu(x))}_\F&=L_{\sigma\mu}-T_{\sigma\mu}u_{GZ}(p^2)\label{uGZ}\ ,
\end{align}
\end{subequations}
satisfy the criteria
\be\label{confGZ}
f_{GZ}(0)=0\qquad\text{and}\qquad u_{GZ}(0)=-1\ .
\ee
However, in this case of a spontaneously broken BRST symmetry it is not entirely clear that
\begin{multline}\label{quartet}
0=\vev{s(A_{\sigma}^a(y)\, \chi^b_\mu(x))} \\
    =\vev{A_{\sigma}^a(y)\, s(\chi^b_\mu(x))}+\vev{(D_\sigma c)^a(y)\, \chi^b_\mu(x)}\ 
\end{multline}
holds, which would imply that only (unphysical) quartet states contribute to  the matrix element of
 \eqref{uGZ}. Due to the equations of motion of the anti-ghost $\bar c$ and of the NL field 
 $b^a$ the longitudinal part of \eqref{quartet} is satisfied.  Although a proof is lacking, 
 it therefore is at least plausible that the transverse part of \eqref{quartet} also holds.
  
The GZ action incorporates non-perturbative features and in fact satisfies the 
criteria (\ref{confGZ}) already at tree-level. 
Expanding the gluonic QEoM~(\ref{gz_eom2}) to tree level yields, 
\begin{align}
 \delta^{ab}\delta_{\sigma\mu}& = \vev{A_\sigma^b(y) \var[S_{GZ}]{A_\mu^a(x)}}_{FT} 
 \label{gz_eom_tl} \\
  & \approx p^2 T_{\mu\nu} \vev{A_\sigma^b(y)\, A_\nu^a(x)}_{FT} +  \vev{A_\sigma^b(y) \, 
  i\partial_\mu b^a(x)}_{FT} \nonumber \\
  & + \gamma^{1/2} g f^{acd} \vev{A_\sigma^b(y) \, (\vph^{c}_{\mu d} -  
  \bar \vph^{c}_{\mu d})(x)}_{FT} \,. \nonumber
\end{align} 
We again have  that the longitudinal part of the gluon propagator is saturated by  the NL
field as in the foregoing investigations. The transverse part of \eqref{gz_eom_tl} is satisfied 
by the  tree-level propagators, given for example in 
\cite{Vandersickel:2012tz, Gracey:2009mj} (with $\lambda^4 = 2 N_c g^2 \gamma$)
\be
 \vev{A_\sigma^b(y)\, A_\mu^a(x)}_{FT} \approx \delta^{ab} \,T_{\sigma\mu} \,\frac{p^2}{p^4 + 
 \lambda^4} \ee
and,
\be \vev{A_\sigma^b(y) \, (\vph^{c,d}_\mu -  \bar \vph^{c,d}_\mu)(x)}_{FT}  \approx  f^{bcd} 
\,T_{\sigma\nu}\, \frac{2g \gamma^{1/2}}{p^4 + \lambda^4} \,.
\ee
The GZ gluon propagator vanishes in the infrared, and $f_{GZ}(0)=0$.  In
addition the last term in \eqref{gz_eom_tl}, derived entirely from the BRST exact term in 
\eqref{gz_eom2}, saturates the  transverse part of the gluonic QEoM at tree-level for 
vanishing momenta,
\be \delta^{ab}T_{\sigma\mu} =  \delta^{ab}T_{\sigma\mu} \frac{p^4}{p^4+\lambda^4} + 
\delta^{ab}T_{\sigma\mu} \frac{\lambda^4}{p^4+\lambda^4} \,.
\ee
Quite strikingly, both criteria of \eqref{confGZ} for a confining phase are thus satisfied  by the
GZ Lagrangian already at tree level. Perturbative calculations to one- and two-loop order in 3  
\cite{Gracey:2010df} and 4 \cite{Ford:2009ar,Gracey:2009zz,Gracey:2009mj} Euclidean dimensions as
well as a non-perturbative infrared analysis~\cite{Huber:2009tx} of  the GZ action show that in the infrared the gluon propagator remains suppressed and the 
ghost propagator diverges more strongly than a massless pole \cite{Zwanziger:1992qr, Huber:2009tx}. This infrared behavior agrees with the original Kugo-Ojima scenario \cite{Kugo:1995km}.  

\section{Conclusion\label{sec_concl}}
In summary, we have formulated as confinement criterion that the gluonic QEoM be saturated by unphysical states in the infrared. In the Higgs and Coulomb phases this is not the case. These conditions thus are sufficient for distinguishing a color confining phase from a Higgs and a
Coulomb phase in linear covariant (LCG) , generalized linear covariant(GLCG), maximal Abelian (MAG), and Coulomb (CG) gauges.  Although the details depend somewhat on the chosen gauge, a universal qualitative criterion emerges in theories with an unbroken BRST or equivariant BRST symmetry that distinguishes between physical and unphysical states. 

In the considered gauges the QEoM of the gauge boson propagator is of the form
\be\label{QEoMG}
 \delta_{\sigma\mu}\delta^{ab} = -\vev{A^a_\sigma(y)\, \partial_\nu F^b_{\nu\mu}(x)}_\F -\vev{A^a_\sigma(y)\, \tilde j^b_\mu (x)}_\F \,,
\ee
where the local current $\tilde j_\mu^a(x)$ differs from the canonical Noether current $j^b_\mu (x)$ of the model by a BRST-exact contribution only,
\be\label{tj}
\tilde
j_\mu^a(x)=j_\mu^a(x) +s \xi_\mu ^a(x) \ .
\ee
In models with unbroken BRST-symmetry, $\tilde j_\mu^a$ thus is physically equivalent to the conserved Noether current $j_\mu^a$.  The criteria distinguish the phases depending upon which term on the right hand side of \eqref{QEoMG} saturates the unity on the left in the infrared.

\begin{widetext}
For the models we considered, the generalized color current $\tilde j_\mu^a(x)$ is given by,
\begin{subequations}
\begin{align}
\tilde j_\mu(x) &={j_\mu}^{\hspace{-.4em}\uone}-i\partial_\mu b &\text{linear covariant  Abelian U(1) (\eqref{qed_dse})},\\
\tilde j_\mu^a(x)&=j^{LCG\,a}_\mu -i s (D_\mu \bar c)^a & \text{in LCG (\eqref{ko_dse})},\\
\tilde j_\mu^a(x)&=j^{GLCG\, a}_\mu -i s_\alpha (D_\mu \bar c)^a & \text{in GLCG (\eqref{cc_dse})},\\
\tilde j_\mu(x)&=j^{MAG}_\mu -i \partial_\mu b & 
\text{SU(2) in MAG  (\eqref{mag_dse})},\\
\tilde j_k^a(x)&=j^{C\, a}_k -i s (D_k \bar c)^a & \text{spatial components in Coulomb gauge (\eqref{coul_spatialeom})},\\
\tilde j_\mu^a(x)&=j^{GZ\, a}_\mu -i s\chi_\mu^a & \text{in GZ (\eqref{gz_dse})}\label{gzdef}.
\end{align}
\end{subequations}
\end{widetext}
In all gauges with unbroken BRST or equivariant BRST symmetry $\tilde j_\mu(x)$ is physically equivalent to the conserved current $j_\mu(x)$ because the additional terms either are BRST-exact or vanish on the physical Hilbert space due to subsidiary conditions. 
 
In the Coulomb phase, the first matrix element in \eqref{QEoMG},   $f(p^2) \delta^{ab}T_{\sigma\mu } =- \vev{A^a_\sigma(y)\, \partial_\nu F^b_{\nu\mu}(x)}_\F$, does not vanish in the infrared. $f(0)\neq 0$ implies the existence of a massless vector boson.

In the Higgs and color-confining phases on the other hand, $f(0)=0$ and the current matrix element saturates \eqref{QEoMG} in the infrared limit $p^2\rightarrow 0$. 

Since no gauge-invariant order parameter discriminates between the Higgs and color-confining phases
\cite{Fradkin:1978dv}, the question arises whether one can distinguish between them at all. All
physical states are colorless in both phases \cite{'tHooft:1979bi,Frohlich:1981yi}.  However,  the states contributing to
$\vev{A^a_\sigma(y)\, \tilde j^b_\mu (x)}_\F$ at low momentum differ in these two phases. In the
confining phase the infrared limit of the transverse part of this current matrix element is \emph{entirely
saturated by unphysical states}. One in  particular can be sure that the phase is confining if a BRST-exact part of  $\tilde j^a_\mu(x)$ saturates the current matrix element. In the
Higgs phase physical states contribute to this matrix element.

We found that such criteria distinguishing the confining from the Higgs and Coulomb phases also exist for Abelian gauge theories.  In the Abelian Coulomb phase, discussed in Sec.~\ref{sec_qed}, the current does not saturate the QEoM of the photon propagator at low momenta and the photon is massless. At tree level in the Abelian Higgs model of Sec.~\ref{sec_higgs}, $f(0)=0$, and the massive physical vector boson saturates the transverse part of the current matrix element. The only BRST-exact contribution to the generalized current in this case is longitudinal.   The SU(2) gauge theory in MAG, discussed in Sec.~\ref{sec_mag}, can be viewed as an Abelian \uone \ gauge theory in a confining phase. If unphysical states created by the $[\se,\seb] A_\mu$-part of the Noether current saturate the current matrix element at vanishing momentum, the theory describes a confining phase.  Only the mutually commuting color charges of the Cartan subgoup are conserved in non-Abelian gauge theories in MAG  and an \emph{
unphysical} part of the corresponding Abelian \emph{Noether current} can saturate the gluonic QEoM in the infrared.

The saturation in GLCG, considered in Sect.~\ref{sec_cc}, resembles that in LCG originally discussed by Kugo and Ojima. In these gauges the model confines color if the BRST-exact term, $ -i s_\alpha (D_\mu \bar c)^a$, of $\tilde j_\mu^a$ saturates the gluonic QEoM in the infrared.

In the non-Abelian Coulomb gauge studied in Sec.~\ref{sec_CG}, only unphysical states that do not satisfy Gauss's Law contribute to the temporal part of the gluonic QEoM in all phases (and at all momenta). The temporal part of the gluonic QEoM thus cannot discriminate between phases.  However,  the theory is again confining if the spatial part of the gluonic QEoM in Coulomb gauge  is  saturated by the BRST-exact $s(D_k \bar c)^a$ term of $\tilde j^a_k$. The corresponding confinement criterion of Coulomb gauge was in addition found to be identical to the horizon condition of minimal Coulomb gauge. 

An equivalence between the Kugo-Ojima confinement criterion and the horizon condition of minimal Landau gauge has also been established in \cite{Zwanziger:1992qr, Dudal:2009xh}. The auxiliary fields also contribute to the conserved Noether currents $j_\mu^{GZ\,a}$ of the GZ action (see \eqref{gz_cur}), but the gluonic QEoM retains the form of \eqref{QEoMG} with $\tilde j_\mu^a$ given by \eqref{gzdef}.  In this model the BRST-exact part $s\chi_\mu^a$ of $\tilde j_\mu^a$ saturates the gluonic QEoM at $p^2=0$ already at \emph{tree level}. There is no massless vector boson, and the gluon propagator at low momentum is suppressed. The GZ-theory in this sense satisfies all the confinement criteria for gauge theories with BRST-symmetry. However, at present it is not known how to define a physical Hilbert space in this model with a spontaneously broken BRST symmetry \cite{Maggiore:1993wq,Vandersickel:2012tz, Dudal:2012sb, Capri:2013naa} and one has to prove that the BRST-exact contribution of the generalized current 
does not create physical states.

\bigskip

{\bf Acknowledgements}\\
VM thanks the members of the Rutgers Newark Physics Department for their hospitality, Lorenz von Smekal for drawing his interest to the Kugo-Ojima scenario, Markus Huber for checking some results presented in Sec.~IIIB and the Institut f\"ur Kernphysik at the Technical University Darmstadt for their support. We thank Nat\'alia Alkofer, Jeff Greensite, Markus Huber and Lorenz von Smekal for helpful discussions. This project was supported by the Austrian Science Fund (FWF), Doctoral Program on {\it Hadrons in Vacuum, Nuclei, and Stars} (FWF DK W1203-N16).

\appendix

\section{Notations and Conventions \label{app_conv}}

In this appendix we fix notations and conventions. Throughout this article gauge 
theories in four-dimensional Euclidean spacetime are considered. 

For QED the covariant derivative of any field $\psi$  with electromagnetic charge $g$
is denoted by
\be
D_\mu \psi    = \partial_\mu \psi - i g A_\mu \psi
\ee
where $A_\mu$ is the gauge connection. The corresponding Abelian field strength is
$F_{\mu\nu} = \partial_\mu A_\nu - \partial_\nu A_\mu$,  and the classical Maxwell
Lagrangian is normalized such that $\LA = \frac 1 4 F_{\mu\nu}F_{\mu\nu}$.  

The covariant derivative of a field in the adjoint representation of an $SU(N)$ 
Yang-Mills theory is written as
\be
D_\mu^{ab} \psi^b = \partial_\mu \psi^a + \cp{A_\mu}{\psi}{a}
\ee 
where the cross product is given by the structure constants $f^{abc}$ of the group,
$\cp{\chi}{\psi}{a} = g f^{abc}\chi^b \psi^c$. In Sec.~\ref{sec_mag}, we use the 
adjoint covariant derivative with respect to the gluon field in the Cartan subalgebra, defined as,
\be \D^{ab}_\mu = \delta^{ab}\partial_\mu  + g f^{aib}A_\mu^i\,.\ee  The non-Abelian field strength is defined by the relation 
\be 
g F_{\mu\nu}^a =  i  \com{D_\mu}{D_\nu}^a ,
\ee
and the classical Yang-Mills Lagrangian by 
$\LYM = \frac{1}{4}\,F^a_{\mu\nu}F_{\mu\nu}^a$. 

The Fourier transform of a correlation function $\vev{\O_1(y)\,\O_2(x)}$ is defined as
\begin{multline} \vev{\O_1(y)\,\O_2(x)}_\F \\ =\frac{1 }{(2\pi)^4}\int\!\!d^4(y\!-\!x)\,\,e^{-ip_{\mu}(y-x)_\mu}\vev{\O_1(y)\,\O_2(x)}\,.\end{multline}

We use an equivalent sign $\equiv$ between expressions that differ by terms that vanish when
the classical equations of motion are satisfied and $\approx$ if expressions coincide to
leading order, usually tree level.

\section{Proof of restored BRST symmetries in MAG \label{app_U1}}
Here we prove that the expectation-values of equivariant BRST and anti-BRST variations (given in \eqref{eBRST}) of  
$\uone$-invariant operators  vanish for an $SU(2)$ gauge theory in MAG, {\it cf.} \eqref{invOp},  
\be
 \vev{\delta_x\Op} = \vev{\se \Op} = \vev{\seb \Op} = 0 \text{,   for all   } \Op\in \W\ , \label{B1}
\ee
where \eqref{mag_U1inv} defines the space $\W$ of \uone-invariant operators. 
With mild restrictions on the topology of space-time, i.e.  the Laplace-operator has to have an 
inverse, \eqref{U1Ward} for any $\Op\in \W$ implies that,
\be
\label{charW}
\vev{\Op\ b(x)}=0\ \text{if}\ \Op\in\W\ .
\ee
The  variations of the \uone-gauge field $A_\mu$ satisfy,
\be\label{sAinv}
\delta_x\se A_\mu=\delta_x\cp{B_\mu}{c}{}=0 
\ee
and
\be  
\delta_x \seb A_\mu=\delta_x \cp{B_\mu}{\bar c}{} = 0  \,.
\ee

$\se A_\mu$ and $\seb A_\mu$ thus are local $\uone-$invariant functionals although 
$A_\mu$ is not. Since the product of two \uone-invariant operators is a 
\uone-invariant operator, \eqref{charW} implies that
\begin{subequations}\label{U1Vertices}
\begin{align}
 0 = \vev{b(x)\ \Op \se A_\mu(y)}& = \vev{  \partial_\nu b(x)\,\Op\, \se A_\mu(y)\ }\,,\\
 0 =  \vev{b(x)\ \Op \seb A_\mu(y)}& = \vev{\partial_\nu b(x)\,\Op\, \seb A_\mu(y)\ } \,.
\end{align}
\end{subequations}
Contracting and taking the limit $y\rightarrow x$, \eqref{U1Vertices} shows that the rhs of 
\eqsref{seWard}~and~(\ref{sebWard}) vanish for functionals $\Op\in\W$. We thus have proven 
\eqref{B1}, that is \eqref{invOp}.

\section{The gluonic QEoM of the $A_0^a$ field \label{app_Coulomb}}
In this Appendix we integrate out the $A_0$ field in \eqref{C_temporaleom}
 and show that the equation is saturated by instantaneous contributions only. 
 With the action $\S_C$ given by the Lagrangian \eqref{c_lag}, 
 the QEoM for the $A_0$-field is
\begin{align} 
\delta(x-y) \delta^{ab} & = \vev{A_0^b(y)\,\var[S_C]{A_0^a(x)}} \label{app_QEoM} \\ 
  & = - \vev{A_0^b(y)\,\left(D_i\left(D_i A_0 - \dot A_i\right)\right)^a(x)} \,. \nonumber
\end{align}
We decompose the right-hand side into two terms, so the gluonic QEoM reads
\be \delta(x-y) = I_1(x-y) + I_2(x-y)\,,\ee
where
\begin{align} I_1(x-y)\delta^{ab} & :=  - \vev{ A_0^b(y)\, (D_i^2 A_0)^a(x)} \nonumber \\
I_2(x-y) \delta^{ab} & := \vev{A_0^b(y) \cp{A_i}{\dot A_i}{a}}\ .
\label{I1andI2}
\end{align}
Here we used $\left(D_i \dot A_i\right)^a = \cp{A_i}{\dot A_i}{a}$, which follows from the 
transverse Coulomb gauge condition $\partial_i A_i^a = 0$.

To improve readability in the following we suppress color indices and add them only where
 necessary. We wish to express these expectation values in terms of an integral over the 
 canonical variables and make use of an identity proven in 
 \cite{Cucchieri:2000hv}:
\begin{align}
\vev{ \O(A_i, A_0)} & =  \vev{ \O\Big(A_i, i { \delta \over \delta \rho } \Big) 
\exp(-i \int d^4x \ \rho  A_0) } \Big|_{\rho = 0}
\nonumber \\ 
& =  N \int dE^{\rm tr} dA^{\rm tr} \ \O \Big(A_i^{\rm tr}, i { \delta \over \delta \rho } \Big)
 \label{canonical} \\
& \quad \times \exp \int d^4x ( i E_i^{\rm tr} \dot{A}_i^{\rm tr} - {\cal H} ) \Big|_{ \rho = 0 }
\,, \nonumber
\end{align}
where $\rho$ is a source for $A_0$.  To obtain this formula, one introduces the color-electric 
field $E_i$ by an auxiliary integration, after which one integrates out $A_0$ and the 
longitudinal part of $E_i$.  This takes one from the Faddeev-Popov formula for integrating over 
$\int d^4A = \int d^3A_i dA_0$ to an integration over the canonical variables of the Coulomb gauge, 
${A^{\rm tr}}_i^b$ and ${E^{\rm tr}}_i^b$, which are the 3-dimensionally transverse 
vector potential and chromoelectric field.  In the last formula, the Hamiltonian density 
is given by
\be {\cal H} := \frac{1}{2} (E^2 + B^2)\,,\ee
where
\begin{align}
B_i^a & =  \epsilon_{ijk} [\partial_j {A^{\rm tr}}_k^a +\frac{1}{2} f^{abc} {A^{\rm tr}}_j^b 
{A^{\rm tr}}_k^c],
\nonumber \\
E_i & =  E_i^{\rm tr} - \partial_i \varphi, 
\nonumber  \\
\varphi & =  M^{-1}(\rho_{\rm coul} + \rho),
\end{align}
and $M = - D_i(A^{\rm tr}) \partial_i$ is the Faddeev-Popov operator of Coulomb gauge.  
Here $\rho_{\rm coul} := - \cp{A^{\rm tr}_i} {E^{\rm tr}_i}{}$ is the color-charge density of the 
dynamical degrees of freedom.  If quarks were present we would have 
$\rho_{\rm coul}^a \equiv - \cp{A^{\rm tr}_i} {E^{\rm tr}_i}{a} + g \bar q \gamma_0 t^a q$.  
From identity~(\ref{canonical}) we obtain 
\be
\vev{ f(A_i) A_0(x)}  = \vev{f(A_i) \ (-iK\rho_{\rm coul})(x) } 
\ee
and
\begin{multline}
\label{identity}
\vev{ f(A_i) A_0(x) A_0(y) }   \\ 
= \vev{ f(A_i) \ [ \ K(x, y) - (K \rho_{\rm coul})(x) \ (K \rho_{\rm coul})(y) \ ] }\,,
\end{multline}
etc.,\ where $(K\rho_{\rm coul})(x) \equiv \int d^4y \ K(x, y) \rho_{\rm coul}(y)$, and
the color-Coulomb kernel is given by
\be K(x, y) \equiv [ M^{-1} ( - \partial_i^2) M^{-1}](x,y)\,.\ee

Identity \eqref{identity}, when applied to \eqref{I1andI2}, gives
\begin{align}
I_1(x-y) & =  - \vev{ \left[ D_i^2 K(x,y) - (D_i^2 K\rho_{\rm coul})(x) (K\rho_{\rm coul})(y)
\right]}
\label{I1_1} \\
I_2(x-y) & =  \vev{ \cp{ A_i}{ \dot A_i}{}\!\!(x) \, (-i)(K\rho_{\rm coul})(y) }\,. \label{I2_1}
\end{align}

We next separate the instantaneous and non-instantaneous parts of these expressions.  
The kernel $K(x,y) = K({\bf x}, {\bf y}) \delta(x_0 - y_0)$ is instantaneous, so the first term 
of $I_1$ is purely instantaneous.  The second term of $I_1$ involves the  canonical fields 
$E^{\rm tr}$ and $A^{\rm tr}$ at time $x_0$ and the canonical fields at time $y_0$.  
These are the dynamical degrees of freedom so their correlators are non-instantaneous. 

Keeping only the instantaneous part in \eqref{I1_1} one gets
    \be I_1(x-y) = - \vev{  D_i^2 K(x,y) }\,,\ee
To separate the instantaneous part in $I_2$, we shall express $\dot A_i^{\rm tr}$ in terms of 
the canonical fields $A_i^{\rm tr}$ and $E_i^{\rm tr}$.  For this purpose we use the fact that 
the integral of a derivative vanishes,
\begin{multline}
 0 = \int dE^{\rm tr} dA^{\rm tr} \, \cp{A^{\rm tr}_i(x)}{\var{ E^{\rm tr}_i(x)}}{a} 
 \Biggl[ \int \!d^4z \, K^{de}(y,z) \\
 \times\cp{A^{\rm tr}_j(z)}{E^{\rm tr}_j(z)}{e} \ \exp( \int\! d^4x ( i E_i^{\rm tr} 
 \dot{A}_i^{\rm tr} - {\cal H} ) \Biggr] \,.
\end{multline}
Note that $ \var[E_j^{\rm tr\,a}(z)]{ E_i^{\rm tr\,b}(x)} = \delta_{ij}^{\rm tr}(x-z) 
\delta^{ab}$.  Here $\delta_{ij}^{\rm tr}(x-z)$ is the kernel of the transverse projector 
$\delta_{ij}I - \partial_i (\partial^2)^{-1} \partial_j$.  This gives
\begin{multline}
0 =  \Big\langle \, \cp{A_i(x)}{[ i \dot A_i(x) - G_i(x)]}{a} \\
 \times \,\int\!d^4z \ K^{de}(y,z) \cp{A_j(z)}{E_j(z)}{e}\\
+ g^2 f^{abc} f^{egc}A_i^b(x)  \int dz \ K^{de}(y,z) A_j^g(z) \delta_{ij}^{\rm tr}(x-z )
\Big\rangle,
\end{multline}
where $G_i^c(x) \equiv { \delta \over \delta E_i^c(z)} \int d^4 z \ {\cal H}(z)$.  
The term in $G_i^c(x)$ involves dynamical fields $E^{\rm tr}$ and $A^{\rm tr}$ at time $x_0$ and 
the second factor involves these fields at time $y_0$ so the term in $G(x, y)$ is 
non-instantaneous.  Keeping only the instantaneous parts, and using 
$K^{de}(y, x) = K^{ed}(x, y)$, we obtain the identity
\begin{align}
&\vev{ \cp{A_i}{ \dot A_i}{a}(x)\, (-i) (K\rho_{\rm coul})^d(y)} \\
=& \vev{ g^2 f^{abc}  f^{cge} A_i^b(x)  \int dz \ \delta_{ij}^{\rm tr} (x-z)\,A_j^g(z) 
\ K^{ed}(z, y)}\,, \nonumber
\end{align}
where the left-hand side is $I_2$.  Because of the transverse projector we may write this as
\be I_2 = \vev{ D_i \int\! d^4z \, \delta_{ij}^{\rm tr} (x-z)  D_j K(z, y) } \,.\ee
\eqref{app_QEoM} in operator notation now reads
\begin{align}
\delta(x-y) & = \vev{ - D_i^2 K(x, y) +D_i \ \delta_{ij}^{\rm tr}  D_j K(x, y)] }
\nonumber  \\
& = \vev{ \ - D_i \ \delta_{ij}^{\rm lo}  D_j K(x, y)] }
\nonumber \\
& = \vev{ - D_i \partial_i (\partial^2)^{-1} \partial_j  D_j K(x, y)] }
\nonumber \\
& = \vev{ M (-\partial^2)^{-1} M K(x, y)] }\,,
\end{align}
 where $\delta_{ij}^{\rm lo} = \partial_i (\partial^2)^{-1}\partial_j$, and so, 
 with $K = M^{-1} (-\partial^2) M^{-1}$,
we obtain the identity
\be \delta(x-y) = \delta(x-y)  \,.\ee
We see that once the $A_0$-field has been integrated out, the gluonic QEoM is satisfied 
identically by the instantaneous parts only.

\end{document}